\documentclass[11pt,a4paper]{article}

\usepackage[
  backend=biber,
  style=phys,        
  articletitle=true, 
  biblabel=brackets  
]{biblatex}
\usepackage{amsfonts}
\usepackage{multirow}
\usepackage{color}
\usepackage{graphicx}
\usepackage{float}
\usepackage{amsmath}
\usepackage{authblk}

\usepackage{subcaption}
\usepackage{comment}
\usepackage{adjustbox} 

\usepackage[dvipsnames]{xcolor}
\definecolor{lightblue}{RGB}{80,160,220}

\usepackage[colorlinks=true, linkcolor=Cerulean, citecolor=blue, urlcolor=blue]{hyperref}

\title{\textbf{Vortex clusters and vortex-filled bubbles\\ in a modified Abelian-Higgs model}}
\author[1]{D. Miguélez-Caballero}
\author[1]{S. Navarro-Obregón}
\author[2,3]{Muneto Nitta}
\author[2]{Keisuke Ohashi}

\affil[1]{Departamento de Física Teorica, Atómica y Óptica and Laboratory for Disruptive Interdisciplinary Science (LaDIS), Universidad de Valladolid, 47011 Valladolid, Spain}
\affil[2]{Department of Physics \& Research and Education Center for Natural Sciences,
Keio University, 4-1-1 Hiyoshi, Yokohama, Kanagawa 223-8521, Japan}
\affil[3]{International Institute for Sustainability with Knotted Chiral Meta Matter (WPI-SKCM$^2$), Hiroshima University, 1-3-1 Kagamiyama, Higashi-Hiroshima, Hiroshima 739-8531, Japan}
\date{}

\begin{document}

\maketitle

\begin{abstract}
We investigate a single-component Abelian–Higgs model with an eighth-order scalar potential whose vacuum manifold consists of two disconnected circles. The model supports two species of vortices and domain walls interpolating between the two vacuum components. We find type-1.5-like behavior characterized by nonmonotonic intervortex interactions and bound states of spatially separated vortices. Inner-vacuum vortices form clusters through short-range repulsion and long-range attraction, whereas outer-vacuum vortices form vortex-filled bubbles in which a closed domain wall confines multiple inner-vacuum-vortex cores. We construct the domain wall analytically, derive the asymptotic vortex–wall interactions, and analyze the fluctuation spectra of coincident multivortices to identify their splitting instabilities. 
\end{abstract}

\section{Introduction}

Vortices are among the simplest examples of topological solitons in gauge theories \cite{Abrikosov1957,Nielsen:1973cs,JaffeTaubes1980,MantonSutcliffe2004}. These localized, finite-energy configurations arise naturally in two spatial dimensions and can also be viewed as transverse cross-sections of extended string-like solitons in three spatial dimensions, commonly referred to as cosmic strings \cite{VilenkinShellard1994}. As a result, vortices have attracted considerable interest in both condensed matter physics and cosmology.

A prototypical theory for studying vortex solutions is the Abelian–Higgs model, in which a complex scalar field is minimally coupled to a $U(1)$ gauge field. In its standard formulation, the scalar sector is governed by a quartic symmetry-breaking potential, where the vacuum manifold is a circle $\mathcal{V}  = \mathbb{S}^1$. The model provides a relativistic analog of the Ginzburg–Landau theory of superconductivity in the static limit. In this context, Abrikosov–Nielsen–Olesen vortices arise as localized configurations carrying quantized magnetic flux \cite{Abrikosov1957,Nielsen:1973cs,JaffeTaubes1980}. Their interactions depend on the value of the dimensionless coupling constant $\lambda$: for $\lambda < 1$, corresponding to type-I superconductivity, vortices attract each other, whereas for $\lambda > 1$, corresponding to type-II superconductivity, they repel each other and can form regular structures known as Abrikosov lattices \cite{Abrikosov1957}. The long-range interactions between well-separated vortices have been determined analytically by modelling the vortices as point sources in the linearized theory around the vacuum \cite{Speight1997}. More recently, the short-range interactions have also been characterized analytically using the spectral data of the Jacobi operator associated with the coincident $n$-vortex \cite{SpeightWinyard2025}. An interesting case occurs at $\lambda = 1$, the so-called 
Bogomol'nyi-Prasad-Sommerfield (BPS) limit, where vortices do not interact statically. In this limit, the low-energy dynamics of vortices admits an elegant geometrical description in terms of geodesic motion on the moduli space. The corresponding moduli space was identified with the space of unordered $n$-tuples of points in the plane, $\mathcal{M}_n = \mathbb{C}^n/S_n$, where $S_n$ represents the the permutation group of $n$ elements \cite{Samols1992}. A vortex configuration with topological charge $n$ therefore possesses $2n$ zero modes, corresponding to the positions of the $n$ vortices in the plane \cite{Weinberg1979}.  
BPS vortices preserve a half of supercharges 
in supersymmetric field theories,
in which they are called $1/2$ BPS states 
\cite{DiVecchia:1977nxl,Witten:1978mh,Edelstein:1993bb,Tong:2005un,Eto:2006pg,Shifman:2007ce,Shifman:2009zz,Tong:2008qd}.
These vortices were generalized to BPS non-Abelian vortices in 
$U(N)$ non-Abelian Higgs model 
\cite{Hanany:2003hp,Auzzi:2003fs,Eto:2004rz,Eto:2005yh,Eto:2006cx,Eto:2006uw,Eto:2006db}, see Refs.~\cite{Tong:2005un,Eto:2006pg,Shifman:2007ce,Shifman:2009zz,Tong:2008qd} as a review.
In this case, the moduli space for 
$n$ vortices is given by 
$\mathcal{M}_n \sim (\mathbb{C}\times {\mathbb C}P^{N-1})^n/S_n$ 
\cite{Eto:2005yh,Eto:2006pg}. 
The metric in the well-separated vortex limit can be found in Refs. \cite{Fujimori:2010fk,Eto:2011pj}. 
Away from the BPS limit, vortex scattering has also been investigated numerically for a wide range of values of the self-coupling constant \cite{ShellardRuback1988,MyersRebbiStrilka1992,BachmaierWereszczynski2026,Auzzi:2007wj}.

Beyond the translational zero modes, the linearized spectrum around a vortex can contain more localized internal modes \cite{GoodbandHindmarsh1995}. Recently, the angular dependence of all the internal modes has been determined in Refs. \cite{AlonsoIzquierdoMiguelezCaballero2024,AlonsoIzquierdo2026b}. Their presence introduces additional degrees of freedom into the dynamics of interacting vortices. In particular, the role of internal modes supported by individual vortices has been investigated both within and beyond the BPS limit \cite{KruschReesWinyard2026,Wereszczynski2026}. It was found that these modes can give rise to either attractive or repulsive forces thereby significantly modifying the resulting vortex dynamics. For example, it has been shown that they can lead to the spectral wall phenomenon \cite{AlonsoIzquierdo2024b,AlonsoIzquierdo2025,AlonsoIzquierdo2026} or to a chaotic scatterings with multiple bounces \cite{Wereszczynski2026,KruschReesWinyard2024}.

An established example beyond the conventional type-I/type-II dichotomy is type-1.5 superconductivity in two-component systems \cite{Babaev2005,PhysRevLett.102.117001,Babaev2010,PhysRevB.81.020506,PhysRevB.83.020503,Babaev2011,Babaev2012,PhysRevB.89.094504}. In these systems, coupled condensate-amplitude fluctuations introduce distinct characteristic length scales in addition to the magnetic penetration depth. In suitable parameter regimes, their interplay produces long-range intervortex attraction and short-range repulsion, allowing vortices to form bound clusters. Under an applied magnetic field, this can lead to a semi-Meissner state in which vortex clusters coexist with vortex-free superconducting regions. Type-1.5 superconductivity thus provides a well-established mechanism for nonmonotonic vortex interactions arising from multiple condensates.

Similar nonmonotonic interactions have also been reported in single-component systems described by Chern--Simons--Landau--Ginzburg theory \cite{Leask2026}, in which vortices carry both electric charge and magnetic flux, or in noncentrosymmetric superconductors, where the magnetic field exhibits an inversion of
the magnetic field at a certain distance from the vortex core \cite{Garaud2020}. Moreover, nonmonotonic vortex interactions have also arisen in single-component systems by modifying the scalar potential while retaining a Maxwell gauge field.
In particular, Abrikosov–Nielsen–Olesen strings with a Coleman–Weinberg potential exhibit nonmonotonic interaction energies with a barrier at intermediate separations, allowing multiply wound strings to remain metastable even in the type-II regime \cite{Eto:2022hyt}. This behavior, termed type-$\overline{1.5}$, illustrates how the nonlinear structure of the scalar potential can modify short-range vortex interactions without introducing additional condensates. Recently, an Abelian-Higgs model with a sixth-order self-interaction potential has been investigated in which both type-1.5 and type-$\overline{1.5}$ regimes are found, depending on the values of the model parameters \cite{Alejandro:2026}. Here, we investigate a different realization based on two disconnected components of the vacuum manifold, supporting bound states of separated vortices and composite configurations involving domain walls.

An interesting modification that can have important consequences for the structure and properties of solitons is to consider theories in which the vacuum manifold consists of several disconnected components. A particularly well-known example is provided by the Chern–Simons–Higgs theory, whose main difference from the Abelian–Higgs model is the replacement of the Maxwell term by a Chern–Simons term \cite{Dunne:1998qy,HorvathyZhang2009}. Requiring the existence of a BPS limit naturally leads to a sixth-order scalar potential with vacuum manifold $\mathcal{V}=\{0\}\cup\mathbb{S}^1$ \cite{JackiwWeinberg1990,HongKimPac1990}. This vacuum structure allows for two distinct classes of vortex solutions: topological and non-topological vortices \cite{HorvathyZhang2009}. Topological vortices are analogous to those of the standard Abelian–Higgs model, carrying quantized magnetic flux, but they also carry electric charge as a consequence of the Chern–Simons term. As a result, they can exhibit anyonic properties, which has motivated their application to models of anyonic superconductivity \cite{KimKimLee1995}. Non-topological vortices, on the other hand, approach the symmetric vacuum $\phi=0$ at spatial infinity and are therefore not characterized by a winding number. Interestingly, their magnetic flux is not topologically quantized and can consequently vary continuously. The interactions between Chern–Simons vortices have also been investigated, with the interaction energy exhibiting both attractive and repulsive behavior depending on the model parameters \cite{Arthur1995}. In the self-dual Chern–Simons–Higgs model, the electric charge carried by the vortices gives rise to velocity-dependent interactions, including a Lorentz-type force \cite{KimLee1994,Dziarmaga1995}. Numerical studies have shown that these interactions lead to nontrivial low-energy scattering dynamics \cite{Strilka2012}. As a consequence, their scattering differs from the familiar $90^\circ$ scattering of standard Abelian–Higgs vortices \cite{Samols1992}.

A natural generalization is therefore to consider a model whose vacuum manifold consists of two disconnected continuous components. In this paper, we achieve this by introducing an eighth-order scalar potential. Potentials of this type have been extensively studied in a variety of contexts. In $(1+1)$-dimensional scalar field theories, they provide a setting in which long-range tail kink solutions can arise \cite{Lohe1979,Manton2019,BazeiaCamposMohammadi2023}. They also occur naturally in the description of phase transitions in condensed matter and cosmological systems, where they can capture the interplay between first- and second-order phase transitions \cite{KhareChristovSaxena2014}. In particular, eighth-order terms in the expansion of the order parameter have been used to describe the phase diagrams of certain ferroelectric materials \cite{Wang2007}, while higher-order corrections have also been found to play an important role in phase transitions associated with ${}^{3}P_{2}$ superfluidity 
of neutrons in neutron stars
\cite{Yasui2019}, admitting 
domain walls 
\cite{Yasui:2019vci},
surface defects \cite{Yasui:2019pgb}, 
and fractional non-Abelian vortices \cite{Kobayashi:2022moc,Kobayashi:2022dae}. Higher-order scalar potentials have been also considered in gauge theories supporting vortex solutions. For instance, when coupled to Einstein gravity, the Chern–Simons–Higgs model admits a Bogomolny limit for both the matter and gravitational sectors for a specific choice of eighth-order scalar potential \cite{London1995,Clement1996,Chung2001}. 

In this paper, we show that a single complex scalar field coupled to a Maxwell gauge field can exhibit type-1.5-like behavior, with nonmonotonic intervortex interactions supporting bound states of spatially separated vortices. The two disconnected vacuum components give rise to two species of vortices, which exhibit distinct binding mechanisms. Inner-vacuum vortices form clusters through a balance between short-range repulsion and long-range attraction, whereas outer-vacuum vortices form vortex-filled bubbles in which a closed domain wall confines multiple inner-vortex cores. We construct the interpolating domain wall analytically and derive the asymptotic vortex–wall interactions, finding repulsion for inner-vacuum vortices and attraction for outer-vacuum vortices. We further analyze the fluctuation spectra of coincident multivortices and identify the splitting instabilities associated with the formation of these bound configurations. These results demonstrate how multiple broken vacua enrich vortex interactions and support composite solitons within a single-component Abelian–Higgs theory.

The paper is organized as follows. First, in Sec. \ref{s:Model}, we introduce the model and identify two species of vortices: inner-vacuum vortices, which interpolate between the symmetric configuration and the inner vacuum, and outer-vacuum vortices, which interpolate between the symmetric configuration and the outer vacuum. Then, in Sec. \ref{s:Interactions_Inner}, we investigate the interaction energy between two inner-vacuum vortices. Apart from the type-I or type-II superconductivity, we find a type-1.5-like regime characterized by a nonmonotonic interaction energy with a minimum at a finite separation. These results indicate the existence of stable vortex clusters. Sec. \ref{s:Vortex_DW} is devoted to constructing the domain wall solution and investigating its interactions with both species of vortices. These results provide insight into the interaction energy between outer-vacuum vortices, which, in certain parameter regimes, can be described as inner-vacuum vortices surrounded by domain walls. In particular, we show that the interaction energy between outer-vacuum vortices can also exhibit type-1.5-like behavior, leading to the formation of vortex-filled bubbles. Finally, Sec. \ref{s:LinearProblem} is devoted to analyze the internal structure around the different species of vortices. As we will show, there are strong differences with respect to the internal structure of vortices from the standard Abelian-Higgs model. Our main conclusions are summarized in Sec.~\ref{s:Conclusions}.

\section{The model and vortices}\label{s:Model}

Let us consider the Abelian-Higgs model, which describes the minimal coupling between a $U(1)$ gauge field $A_{\mu}(\vec{x},t) = (A_0(\vec{x},t),A_1(\vec{x},t),A_2(\vec{x},t))$ and a complex scalar field $\phi(\vec{x},t) = \phi_1(\vec{x},t) + i\, \phi_2(\vec{x},t)$ in $2 + 1$ dimensions
\begin{equation}
    \tilde{\mathcal{L}}=-\frac{1}{4} F_{\mu\nu}F^{\mu \nu}+\frac{1}{2}\overline{D_\mu \phi} D^\mu \phi - \dfrac{\tilde{\lambda}}{8}\, \tilde{U}_{\sigma}(|\phi|^2)\,,
\end{equation}
where the Maxwell tensor is given by $F_{\mu\nu} = \partial_{\mu}A_{\nu} - \partial_{\nu}A_{\mu}$ and the covariant derivative by $D_i = \partial_i - i e A_i$. We will consider that the potential $\tilde{U}_{\sigma}(|\phi|^2)$ is defined as 
\begin{equation}\label{eq:Potential_Old}
    \tilde{U}_{\sigma}(|\phi|^2) =
    \dfrac{1}{\sigma^4}
    \left(\nu^2-|\phi|^2\right)^2
    \left(\sigma^2-|\phi|^2\right)^2\,.
\end{equation}
It is convenient to rescale the coordinates, the fields and the parameters as follows
\begin{equation}
    x^{\mu} \rightarrow x^{\mu}/(q \nu)\,, \quad     \phi \rightarrow \nu \phi\,, \quad A_{\mu} \rightarrow \nu A_{\mu}\,, \quad \sigma \rightarrow \nu \sigma\,.
\end{equation}
In terms of these rescaled fields, the model reads 
\begin{equation}\label{eq:Lagrangian}
    \mathcal{L}= \dfrac{\nu}{e}\left(-\frac{1}{4} F_{\mu\nu}F^{\mu \nu}+ \frac{1}{2}\overline{D_\mu \phi} D^\mu \phi - \dfrac{\lambda}{8}\, U_{\sigma}(|\phi|^2)\right)\,,
\end{equation}
with
\begin{equation}\label{eq:Potential}
    U_{\sigma}(|\phi|^2) =
    \dfrac{1}{\sigma^4}
    \left(1-|\phi|^2\right)^2
    \left(\sigma^2-|\phi|^2\right)^2\,,
\end{equation}
and the coupling constant is $\lambda = \tilde{\lambda}/e^2$. In terms of the rescaled field, the covariant derivative is defined as $D_\mu = \partial_\mu - i A_\mu$. In these coordinates, we have effectively set $e = 1$ and $\nu = 1$. From now on, we will restrict ourselves to the range $\sigma \geq 1$. Note that the $\sigma \to \infty$ limit continuously reduces the potential to the standard $\phi^4$ form.

\begin{figure}[ht!]
    \centering
    \includegraphics[width=0.55\linewidth]{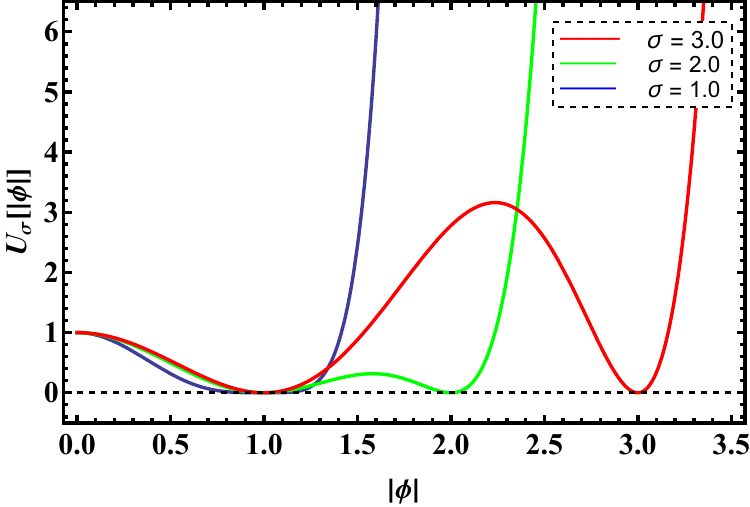}
    \caption{\small Scalar potential \eqref{eq:Potential} as a function of the scalar field modulus. For $\sigma \neq 1$ the potential exhibits two degenerate continuous vacuum manifolds located at $|\phi| = 1.0$ and $|\phi| = \sigma$. For $\sigma = 1$ the two vacuum manifolds coincide.}
    \label{fig:Potential}
\end{figure}

A plot of the potential $\eqref{eq:Potential}$ is shown in Fig. \ref{fig:Potential} for different values of $\sigma$. This potential admits an inner set of degenerate vacua located at $\mathcal{V}_1 = \{\phi:|\phi|^2 = 1\}$ and an outer set of degenerate vacua located at $\mathcal{V}_2 = \{\phi:|\phi|^2 = \sigma^2\}$. Hence, the vacuum manifold of the potential $\eqref{eq:Potential}$ consists in the union of two circles, $\mathcal{V} = \mathcal{V}_1\cup\mathcal{V}_2  \cong \mathbb{S}^1\cup\mathbb{S}^1$. Here, the topological sectors are labelled by $(v,n)$, where $v \in \{1,\sigma\}$ denotes the vacuum component and $n \in \pi_1( \mathcal{V}_i) \simeq \mathbb{Z}$ accounts for the winding number. Therefore, this theory exhibits two types of local vortex-like configurations, distinguished by the asymptotic vacuum approached by the scalar field, namely $\mathcal{V}_1$ and $\mathcal{V}_2$. Both types are characterized by an integer winding number $n$, and can thus be regarded as belonging to the $(1,n)$ and $(\sigma,n)$ sectors, respectively. In addition to vortex configurations, the disconnected structure of $\mathcal{V}$ allows for domain-wall configurations interpolating between $\mathcal{V}_1$ and $\mathcal{V}_2$. Such configurations connect a point on the inner vacuum circle to a point on the outer one. The asymptotic phases do not affect the domain-wall topological sector, since points with different phases on the same vacuum circle belong to the same connected component of the vacuum manifold. For the limiting value $\sigma = 1$, the two vacuum manifolds coincide, and the vacuum manifold reduces to $\mathcal{V} \cong \mathbb{S}^1$. As a consequence, the domain wall ceases to exist and the two kinds of vortices degenerate into a single vortex.

Imposing  the temporal gauge $A_0 = 0$, the Euler-Lagrange equations governing the dynamics of this model read
\begin{equation}\label{Eq:EquationSecondOrdenCartesianCoordinates}
    \begin{split}
         \partial_{0}^2\phi  &=D_1D_1\phi+D_2D_2\phi - \dfrac{\lambda}{4} \dfrac{\partial U_{\sigma}}{\partial \overline{\phi}},\\
       \partial_{0}^2A_1  &=\partial_{2}\partial_2A_1-\partial_{1}\partial_2A_2-\frac{i}{2}\left(\overline{\phi} D_1\phi-\phi \overline{D_1\phi}\right),\\
        \partial_{0}^2A_2 &=\partial_{1}\partial_1A_2-\partial_{1}\partial_2A_1-\frac{i}{2}\left(\overline{\phi} D_2\phi-\phi \overline{D_2\phi}\right).
    \end{split}
\end{equation}
For a static field configuration, the corresponding energy functional related to \eqref{eq:Lagrangian} can be written as
\begin{equation}
\label{eq:staticenergy}
    V=\frac{1}{2}\int_{\mathbb{R}^2}\left[ B^2+\overline{D_1\phi}D_1 \Phi + \overline{D_2\phi}D_2\Phi + \dfrac{\lambda}{4} U_{\sigma}(|\phi|^2) \right]dx_1 dx_2,
\end{equation}
where $B:=F_{1 2}=\partial_1A_2-\partial_2 A_1$ is the magnetic field.

\begin{figure}[ht!]
    \centering
    \begin{minipage}{\linewidth}
        \centering
        \includegraphics[width=0.9\linewidth]{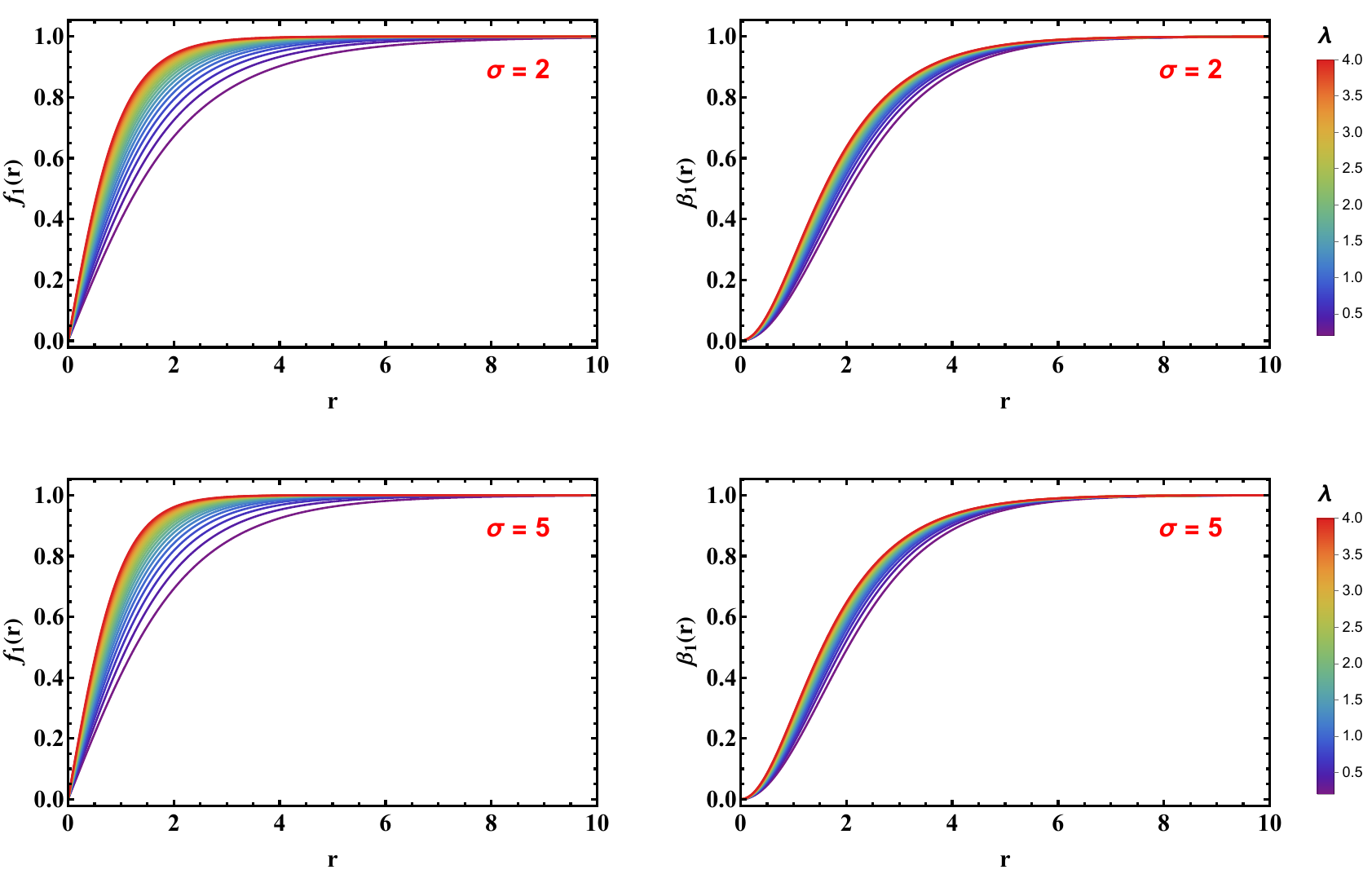}
    \end{minipage}
    
    \caption{\small Scalar (left) and vector (right) field profiles of the inner $1$-vortex for different values of the self-coupling constant $\lambda$.  The upper and lower panels correspond to $\sigma = 2$ and $\sigma = 5$, respectively.}
    \label{fig:Profiles_Inner}
\end{figure}

\begin{figure}[ht!]
    \centering
    \begin{minipage}{\linewidth}
        \centering
        \includegraphics[width=0.9\linewidth]{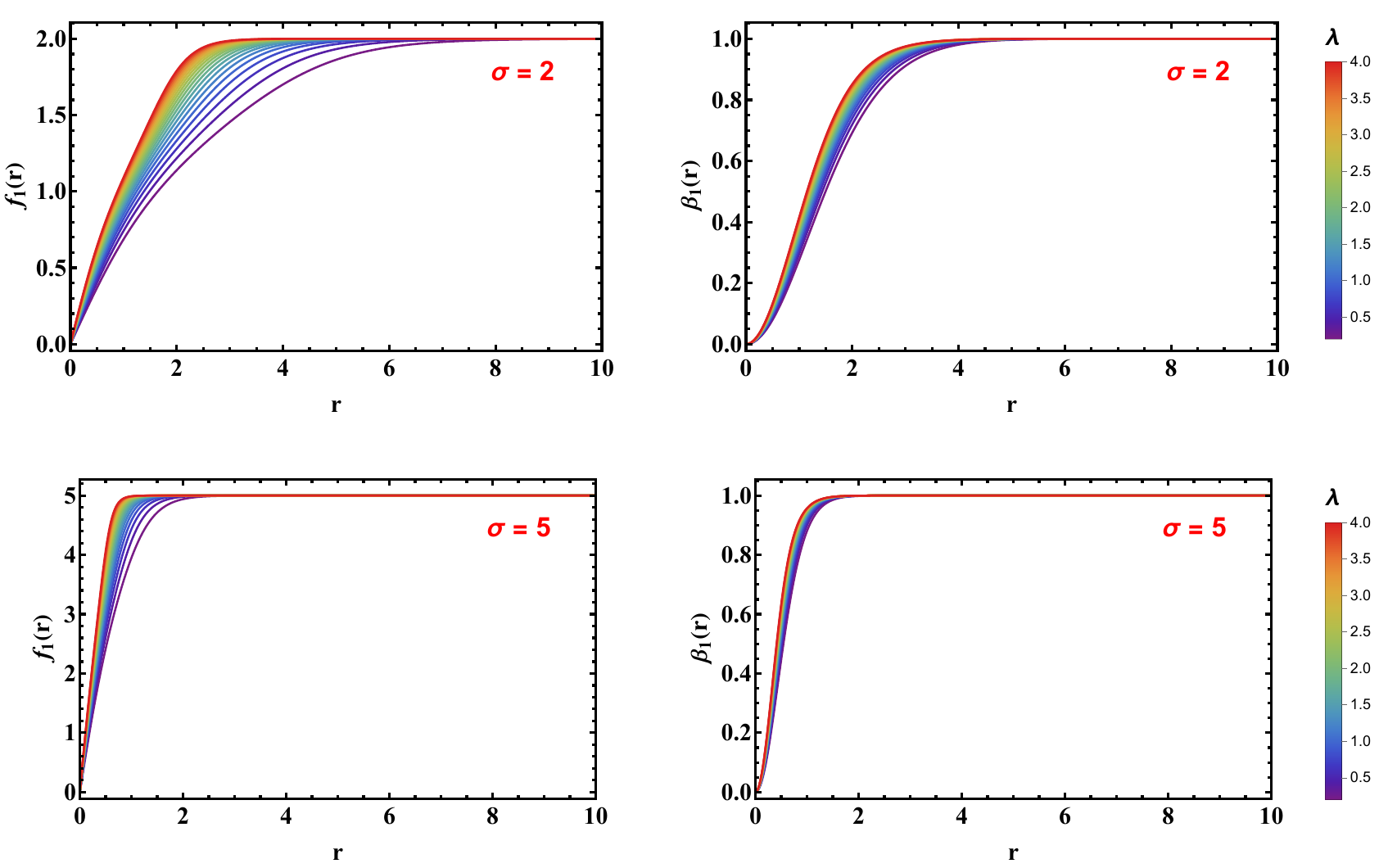}
    \end{minipage}
    \caption{\small Scalar (left) and vector (right) field profiles of the outer $1$-vortex for different values of the self-coupling constant $\lambda$. The upper and lower panels correspond to $\sigma = 2$ and $\sigma = 5$, respectively.}
    \label{fig:Profiles_Outer}
\end{figure}

\begin{figure}[ht!]
    \centering
    \begin{minipage}{\linewidth}
        \centering
        \includegraphics[width=0.87\linewidth]{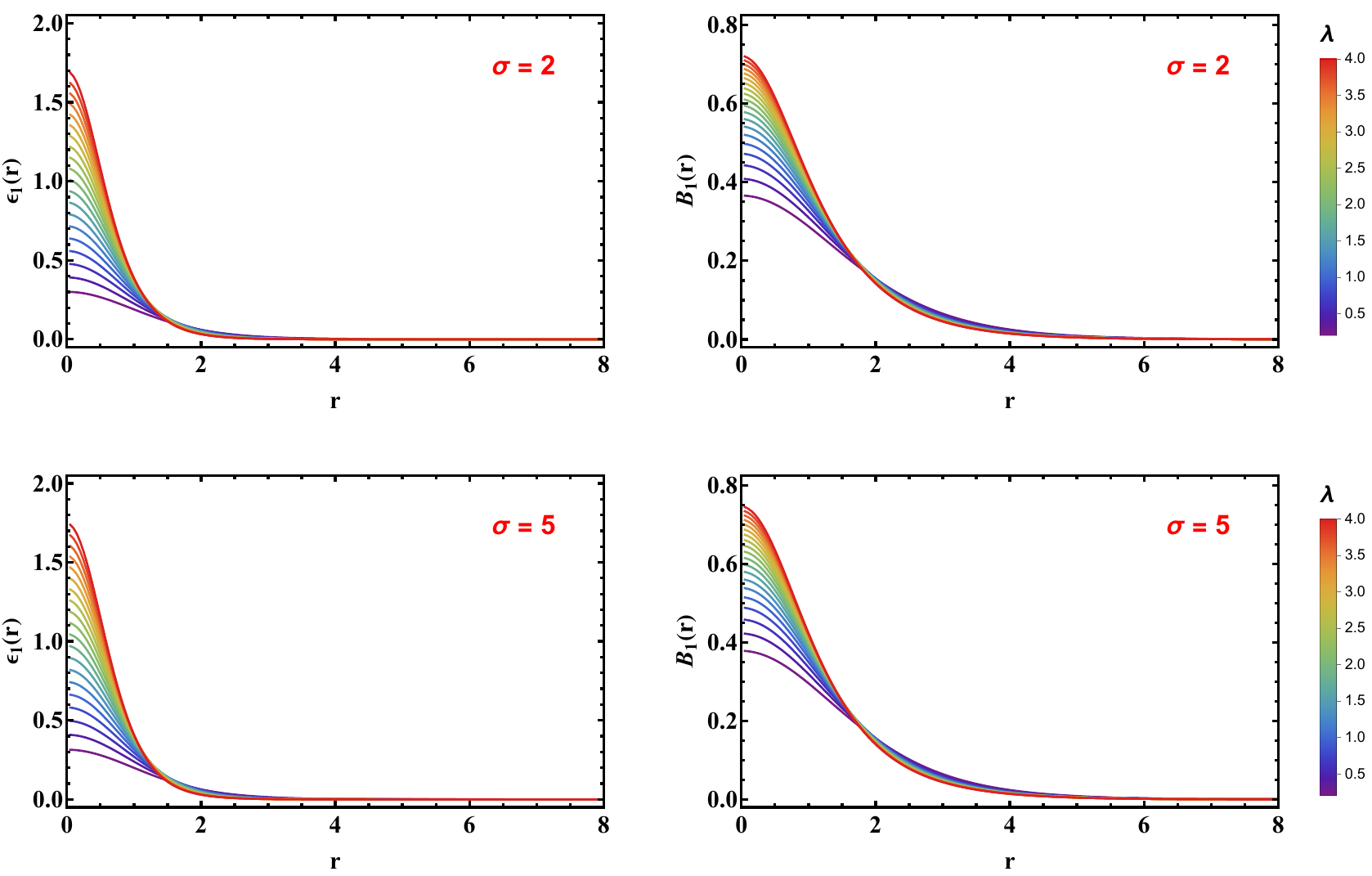}
    \end{minipage}
    
    \caption{\small Energy density (left) and magnetic field (right) profiles of the inner $1$-vortex for different values of the self-coupling constant $\lambda$. The upper panel shows the results for $\sigma = 2$, while the lower panel shows the results for $\sigma = 5$.}
    \label{fig:Energy_Magnetic_Inner}
\end{figure}

\begin{figure}[ht!]
    \centering
    \begin{minipage}{\linewidth}
        \centering
        \includegraphics[width=0.87\linewidth]{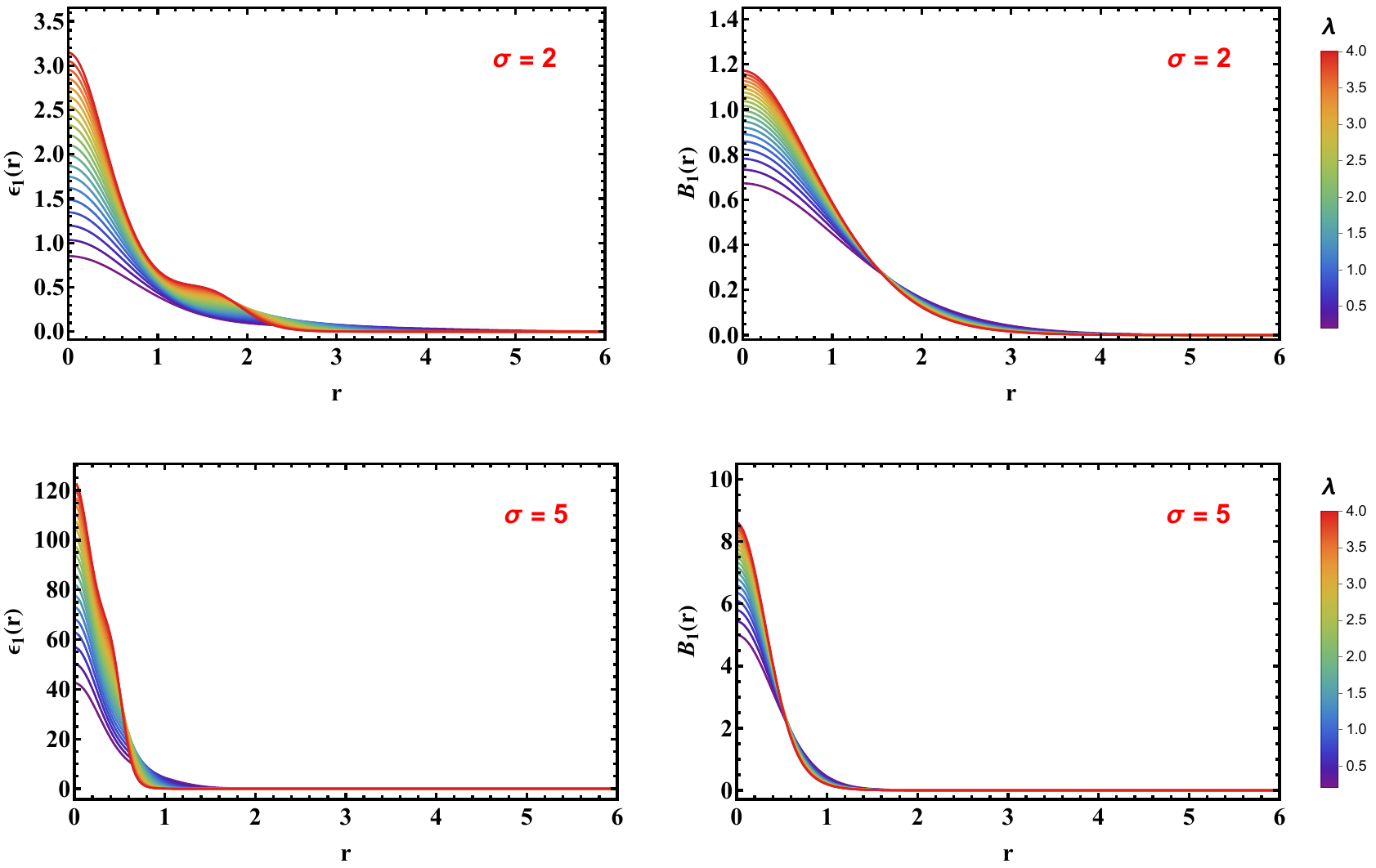}
    \end{minipage}
    
    \caption{\small Energy density (left) and magnetic field (right) profiles of the outer $1$-vortex for different values of the self-coupling constant $\lambda$. The upper panel shows the results for $\sigma = 2$, while the lower panel shows the results for $\sigma = 5$.}
    \label{fig:Energy_Magnetic_Outer}
\end{figure}

In order to determine static axially-symmetric $n$-vortex solutions, let us assume the Nielsen-Olesen ansatz \cite{Nielsen:1973cs}
\begin{equation}\label{e:ANO}
    \phi(r,\theta) = f_n(r)e^{i n \theta}\,, \quad A_r(r,\theta) = 0\,, \quad A_{\theta}(r,\theta) = \dfrac{\beta_n(r)}{r}\,, 
\end{equation}
where the corresponding magnetic field is given by $B = \beta'_n(r)/r$. By introducing this ansatz into the field equations \eqref{Eq:EquationSecondOrdenCartesianCoordinates} we obtain the following set of coupled ordinary differential equations
\begin{equation}\label{Eq:BetaF_Equations}
    \begin{split}
    \dfrac{d^2 f_n}{d r^2}&+\frac{1}{r}\dfrac{d f_n}{dr }-\frac{1}{r^2}(n-\beta_n)^2 f_n + \frac{\lambda}{2\sigma^4}(\sigma^2-f_n^2)(1-f_n^2)(1+\sigma^2-2f_n^2)f_n\,=0,\\
    \dfrac{d^2\beta_n}{d r^2}&-\frac{1}{r}\frac{d \beta_n}{d r}+(n-\beta_n) f_n^2 =0\,.
    \end{split}
\end{equation}
The regularity condition at the origin requires $f_n(0) = \beta_n(0) = 0$, while finiteness of the energy requires $\beta_n(\infty) = n$ and $f_n(\infty)$ to approach a vacuum value $v = 1$ or $v= \sigma$. The vortex profiles are obtained by numerically solving the static radial equations \eqref{Eq:BetaF_Equations} using a gradient flow method. 
Figs.~\ref{fig:Profiles_Inner} and \ref{fig:Profiles_Outer} show the scalar and vector radial profiles of the inner- and outer-vacuum vortices, respectively, as functions of the coupling constant $\lambda$ for different values of $\sigma$. The corresponding static energy densities and magnetic field profiles are shown in Figs. \ref{fig:Energy_Magnetic_Inner} and \ref{fig:Energy_Magnetic_Outer} for the inner and outer-vacuum vortices, respectively. The profiles of the energy density indicate that the outer-vacuum vortex can be realized as an inner-vacuum vortex surrounded by a domain wall for certain regimes. The domain wall solution and its role on the outer-vacuum vortex will be analyzed in more detail in Sec. \ref{s:Vortex_DW}. 

The power expansion of a $n$-vortex around its core shows that
\begin{equation}
\begin{aligned}
    f_n(r) &\sim r^n\sum_{i = 0}^{\infty}d_{2i}r^{2i}\,,\\
    \beta_n(r) &\sim r^2\left(c_0 + \sum_{i = n}^{\infty}c_{2i}r^{2i} \right)\,,
\end{aligned}
    \qquad r\to 0\,,
\end{equation}
whereas at large distances from the core, the profiles behave as
\begin{equation}\label{e:asymptotic}
\begin{aligned}
    f_n(r) &\sim v - \dfrac{q}{2\pi}\,K_0\left(m_H^{(v)}r\right)\,,\\
    \beta_n(r) &\sim n - \dfrac{m}{2\pi}\,r K_1\left(m_g^{(v)}r\right)\,,
\end{aligned}
    \qquad r\to\infty\,,
\end{equation}
where $K_l(r)$ denotes the Bessel function of second kind and order $l$, and where $q$ and $m$ are two unknown coefficients obtained by fitting \eqref{e:asymptotic} to the asymptotic numerical profiles of $f_n(r)$ and $\beta_n(r)$. Strictly speaking, this approximation for the scalar field fails when $m_H^{(v)} > 2m_g^{(v)}$, since nonlinear effects cause the long-range scalar contribution to be replaced by an $\mathcal{O}(e^{-2m_g^{(v)} r})$ decay \cite{Perivolaropoulos1993}. 

Here, the scalar and vector masses at the two vacua are given by
\begin{equation}\label{e:masses}
\begin{aligned}
v=1:&\qquad
m_H^{(1)}=\sqrt{\lambda}\frac{\sigma^2-1}{\sigma^2}\,,
\qquad m_g^{(1)}=1\,,
\\
v=\sigma:&\qquad
m_H^{(\sigma)}=\sqrt{\lambda}\frac{\sigma^2-1}{\sigma}\,,
\qquad m_g^{(\sigma)}=\sigma\,,
\end{aligned}
\end{equation}
Note that, in this model, the ratio of the Higgs mass to the gauge field mass is the same around both vacua. Specifically, $\gamma = m_H^{(v)}/m_g^{(v)} = \sqrt{\lambda}(\sigma^2-1)/\sigma^2$. This parameter dictates the long-range interactions between vortices and will be essential for the discussion below. Finally, remark that for the limit $\sigma = 1$, the mass of the Higgs vanishes. For this regime, we expect the vortex to approach its vacuum value according to a power law, indicating the presence of a long-range tail. However, this problem will not be addressed in the present work.

Having characterized the different types of vortices arising in the model, we now turn our attention to their interactions. Through this analysis we will identify regimes of total attraction or repulsion as well as the emergence of inner-vortex clusters and vortex-filled bubbles.

\section{Inner-vortex clusters}\label{s:Interactions_Inner}

The interaction between solitons is mediated by their overlapping tails. Far from their cores, the fields are well approximated by solutions of the linearized field theory, allowing the solitons to be replaced by effective sources.   

\begin{figure}[ht!]
    \centering
 \includegraphics[trim=0cm 12.cm 0cm 0, clip=true, width=0.98\textwidth]{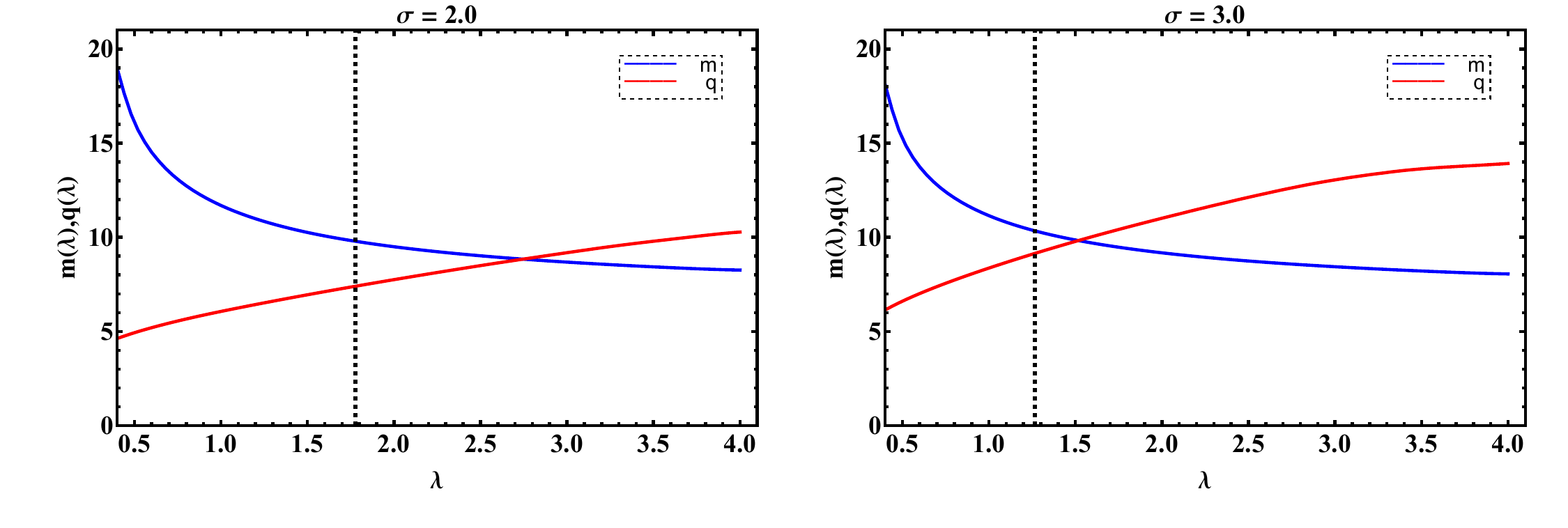}
 
 \hspace{0.05cm} \includegraphics[width=0.47\linewidth]{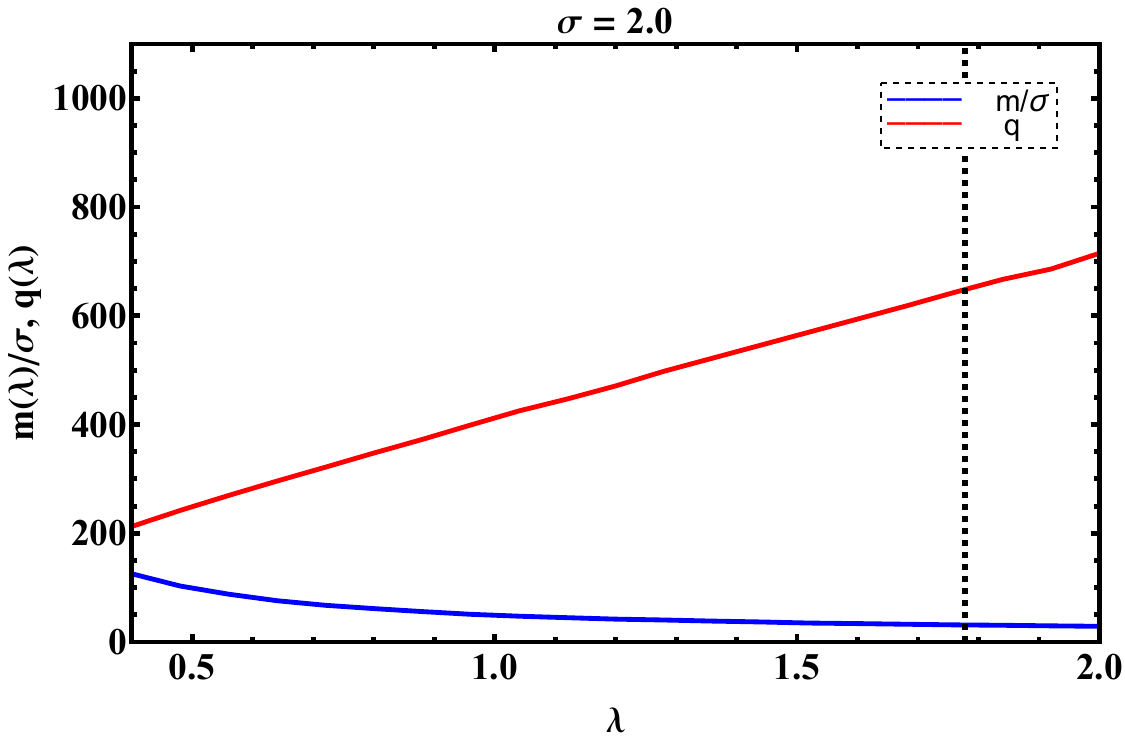}\hspace{0.1cm}
  \includegraphics[width=0.465\linewidth]{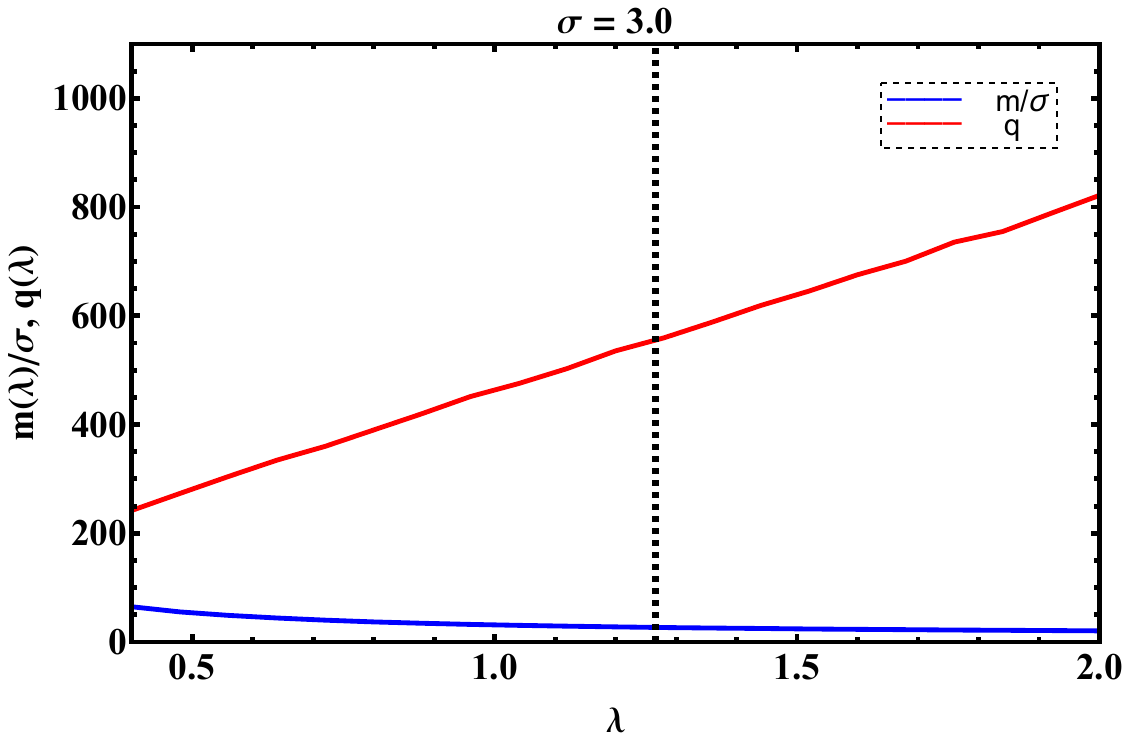}

    \caption{\small Numerical values of the coefficients $m/\sigma$ and $q$, obtained by fitting the numerical profiles of $f_n(r)$ and $\beta_n(r)$ to the analytical asymptotic expression \eqref{e:asymptotic}. We consider $\sigma = 2$ and $\sigma = 3$. The upper panel corresponds to the inner-vacuum vortex, while the lower panel corresponds to the outer-vacuum vortex. The dashed line indicates the value of $\lambda$ for which the Higgs and gauge masses are equal.}
    \label{fig:q_m}
\end{figure}

An asymptotic formula for the interaction energy between
well-separated vortices was derived in Ref.~\cite{Speight1997}.  The main idea is to reproduce the asymptotic behavior of the vortex within the linearized field theory by coupling the fields to a scalar density and a vector current. In this effective description, the vortex is represented by a scalar monopole of charge $q$ and a magnetic dipole of moment $m$. The interaction Lagrangian for a pair of sources then allows us to determine the interaction energy between vortices. 
This general expression can be generally written as
\begin{equation}\label{eq:U_int}
    U_{\text{int}}(d) = \dfrac{1}{2\pi}\left[\left(\dfrac{m}{m_g^{(v)}}\right)^2 K_0(m_g^{(v)} d) - q^2K_0(m_H^{(v)}d)\right]\,,
\end{equation}
where $d$ denotes the separation between vortices. Fig. \ref{fig:q_m}  shows the dependence of the constants $m$ and $q$ on $\lambda$ for inner and outer-vacuum vortices with $\sigma =2$ and $\sigma = 3$. These parameters are obtained by fitting the numerical profiles of the inner-vacuum vortices to the asymptotic expressions given in \eqref{e:asymptotic}. This general  expression shows that the scalar contribution is attractive, while the magnetic contribution is repulsive. Consequently, at sufficiently large separations, the interaction is dominated by the contribution associated with the lighter mass. Thus, when $m_H^{(v)} < m_g^{(v)}$, the scalar contribution dominates and the vortices are expected to attract, whereas when $m_H^{(v)} > m_g^{(v)}$, the magnetic contribution dominates and the vortices are expected to repel. A limiting case occurs when $m_H^{(v)} = m_g^{(v)}$: if, in addition, $q = m/m_{g}^{(v)}$, the two contributions exactly cancel, and the interaction energy vanishes at all separations. This cancellation would suggest a BPS limit. However, these conditions are not simultaneously satisfied in our model, because the Abelian-Higgs model only possess a BPS limit for the standard quartic potential. This is corroborated in Fig. \ref{fig:q_m}, where the dashed line indicates the condition under which the two masses coincide. As we will see later, the relative magnitude of $q$ and $m/m_g^{(v)}$ when $m_H^{(v)} = m_g^{(v)}$ will be a crucial point.

The asymptotic expression \eqref{eq:U_int} is useful for determining whether vortices attract or repel at large separations. However, it provides limited information about the interaction energy at intermediate distances due to nonlinearities. To address this regime, we consider two unit vortices $(\phi^{(1)},A_i^{(1)})$ located at $(x_1,x_2) = (\pm d/2,0)$, respectively. The initial configuration is then constructed by superposing the two vortices according to the Abrikosov ansatz
\begin{align}\label{e:Abrikosov_Vortices}
    \phi^{(2)}(\vec{x}, d) &= \frac{1}{v}\phi^{(1)}(\vec{x}, d/2)\,\phi^{(1)}(\vec{x}, -d/2)\,,\\
    A_i^{(2)}(\vec{x}, d) &= A_i^{(1)}(\vec{x}, d/2) + A_i^{(1)}(\vec{x}, -d/2)\,.
\end{align}
We then minimize the static energy over all degrees of freedom using a local relaxation method
\begin{equation}
    \partial_{\tau}X = - \dfrac{\delta V}{\delta X}\,,
\end{equation}
where $\tau$ is the relaxation time and $X = \{\phi_1,\phi_2,A_i\}$. To keep the vortex cores fixed, we impose the constraints $\phi(\pm d/2,0) = 0$ throughout the evolution. Finally, the interaction energy is defined as $U_{\text{int}}(d) = V_2(d) - 2V_1$, where $V_2(d)$ denotes the minimized energy of the vortex-vortex configuration at separation $d$, and $V_1$ is the static energy of an isolated $1$-vortex solution. We have performed the simulations in a box of size $[-L,L]\times[-L,L]$ with $L = 30$.

\begin{figure}[ht!]
    \centering
    \includegraphics[width=0.98\linewidth]{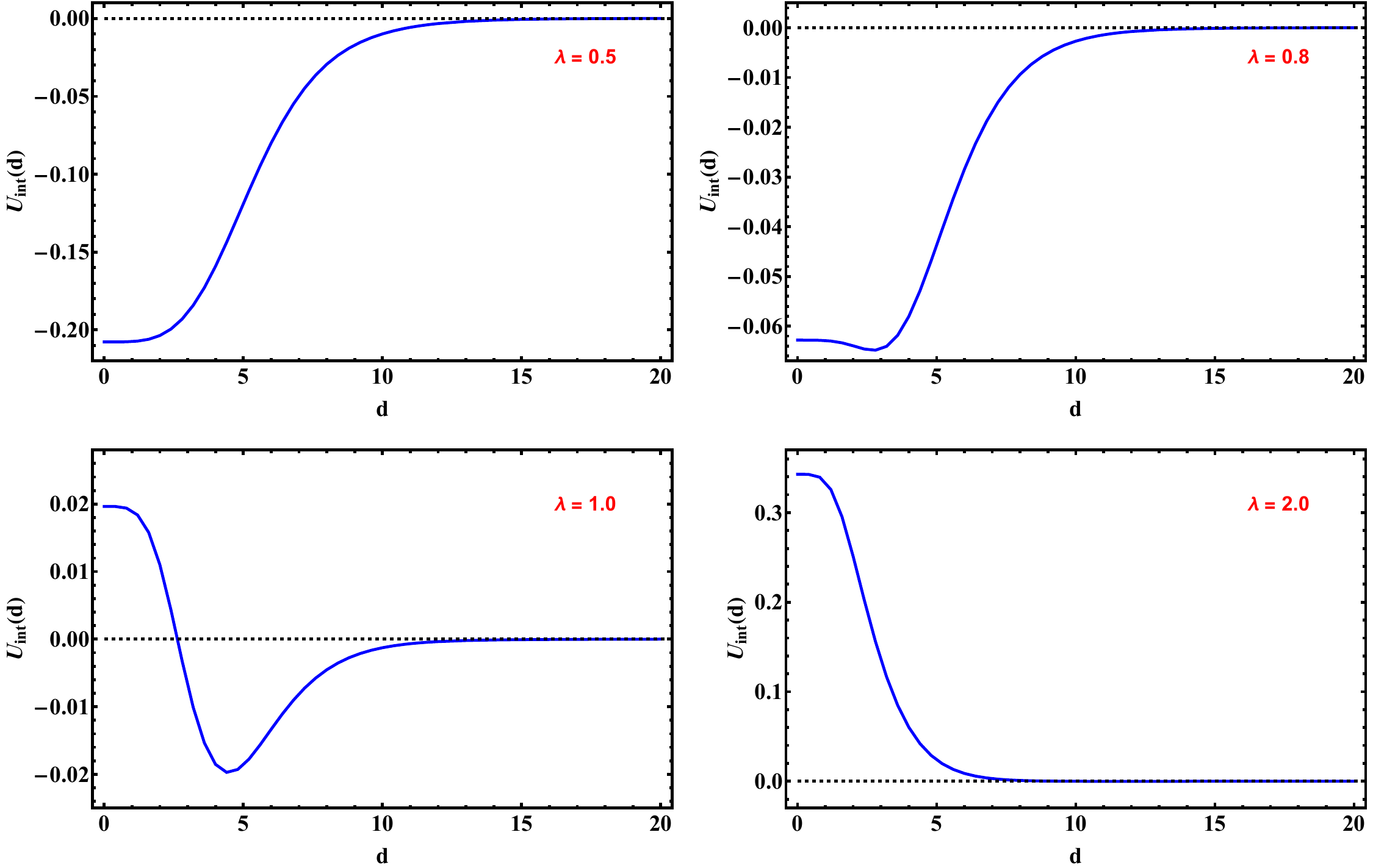}
    
    \caption{\small Interaction energy between two inner $1$-vortices as a function of the separation $d$ for different values of the self-coupling constant $\lambda$ and $\sigma=2$, where the critical values are $\lambda_{c1} \approx 0.75$ and $\lambda_{c2} \approx 1.78$.}
    \label{fig:InteractionEnergy}
\end{figure}

In Fig. \ref{fig:InteractionEnergy} we depict the interaction energy between two inner-vacuum vortices for different values of the self-coupling constant $\lambda$. As a representative example, we consider $\sigma = 2$. The analysis reveals that we can distinguish three different regimes: For $\lambda < \lambda_{c1}$, the interaction between two vortices is purely attractive, as in conventional type-I superconductors, so that the $2$-vortex configuration is stable. For $\lambda_{c1} < \lambda < \lambda_{c2}$, the interaction develops a finite-separation global minimum at $d^*{(\lambda,\sigma)}$, corresponding to a vortex molecule with the two vortices located at $(\pm d^*/2,0)$. Specifically, for values of $\lambda$ just above $\lambda_{c1}$, the coincident $2$-vortex configuration still has lower energy than two well-separated unit vortices, although it is unstable against splitting due to the appearance of the finite-separation minimum. As $\lambda$ is increased further, the short-range repulsive barrier becomes progressively higher, and the energy of the coincident $2$-vortex eventually exceeds that of two well-separated unit vortices. Meanwhile, the global minimum moves towards larger separations and becomes progressively shallower. Finally, for $\lambda > \lambda_{c2}$, the finite-separation minimum disappears and the interaction becomes purely repulsive, as in conventional type-II superconductors. Here, $\lambda_{c1}$ and $\lambda_{c2}$ are considered to be related to the properties of the vortices when the relative separation is small or big. We discuss the emergence of this threshold in Sec. \ref{s:LinearProblem}. On the other hand, the upper critical value $\lambda_{c2}$ is trivially determined exactly by the mass degeneracy condition, i.e., $\gamma = 1$. Therefore, $\lambda_{c2} = \sigma^4/(\sigma^2-1)^2$. Numerically, we have verified that $\lambda_{c1}$ increases while $\lambda_{c2}$ decreases as $\sigma$ increases, both approaching $\lambda_{c1}=\lambda_{c2}=1$ in the limit $\sigma\to\infty$. In this limit, the standard Abelian–Higgs model is recovered, and the critical values corresponds to the BPS limit. 

The nonmonotonic interaction energy observed in our model resembles that found in type-1.5 superconductors, that can arise from the interplay between multiple superconducting components \cite{Babaev2005,Babaev2010,Babaev2011,Babaev2012}. In our $\phi^8$ model, the observed nonmonotonic interaction energy is a direct consequence of higher-order nonlinearities. In the standard Abelian–Higgs model, the Higgs mass $m_H = \sqrt{\lambda}$, which establishes the characteristic length scale of the scalar sector, and the central potential energy density $U_{\phi^4}(0) = \lambda/8$, which measures the energy cost of destroying the condensate, share a rigid parametric dependence on the single self-coupling $\lambda$. As a result, the core stiffness and the asymptotic decay rate cannot be independently varied. In particular, when the Higgs and gauge masses are degenerate, it is well known that $m = q$, the interaction forces cancel exactly, allowing for the existence of BPS states. However, the higher-order nonlinearities introduced in our model modify the curvature of the potential via the parameter $\sigma$. Specifically, it modifies the Higgs mass to $m_H^{(1)} = \sqrt{\lambda}(\sigma^2 - 1)/\sigma^2$, while leaving the core energy density $U_{\sigma}(0) = \lambda/8$ invariant. Consequently, this allows us to decouple the asymptotic tail dynamics from the core energetics. Although the dominant interaction at large distances is still governed by lightest mass, the repulsive magnetic and attractive scalar contributions can compete at short distances, driving the possible emergence of a global minimum at a finite separation between the vortices. This behavior can also be inferred from the $q$ and $m$ coefficients. Here, the $\sigma$-dependence affects the values of $m$ and $q$ and breaks  the relation $m = q$ at the critical value. Specifically, the upper panel of Fig. \ref{fig:q_m} shows that $m > q$ at the critical value $\gamma = 1$, so the vector-mediated interaction dominates at large distances, leading to repulsion. However, slightly below the critical value, the scalar-mediated interaction becomes dominant asymptotically, causing the interaction to turn attractive at sufficiently large distances. Nevertheless, since $m\gg q$, the magnetic repulsion remains strong enough to compete with the scalar attraction over an intermediate range of distances. This implies the emergence of a local minimum $d = d^*$ in the interaction potential. This local minimum is expected to persist even significantly below the critical value.

\begin{figure}[ht!]
    \centering
        \centering
        \includegraphics[width=0.4\linewidth]{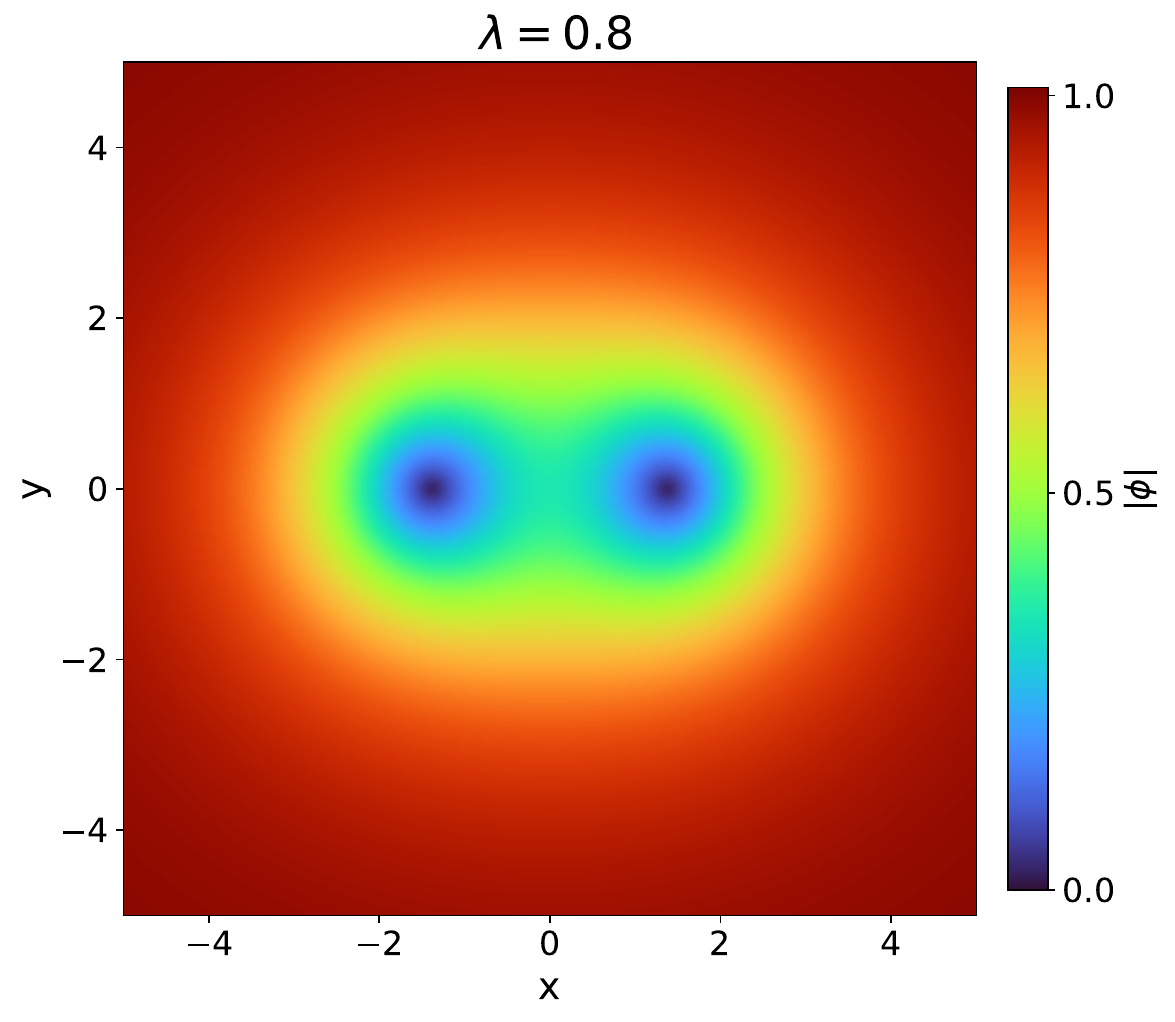}\includegraphics[width=0.4\linewidth]{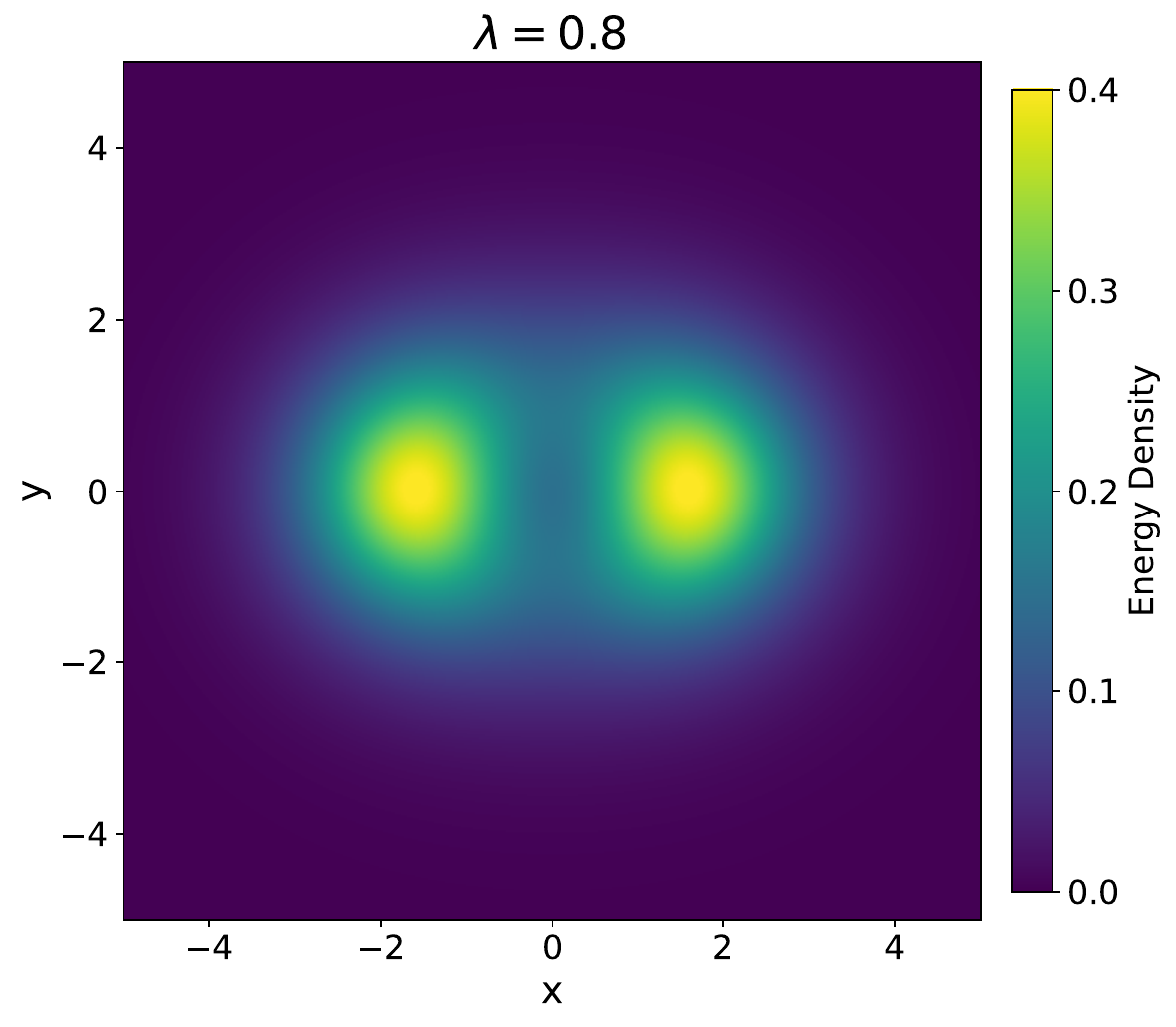}%
        
        \includegraphics[width=0.4\linewidth]{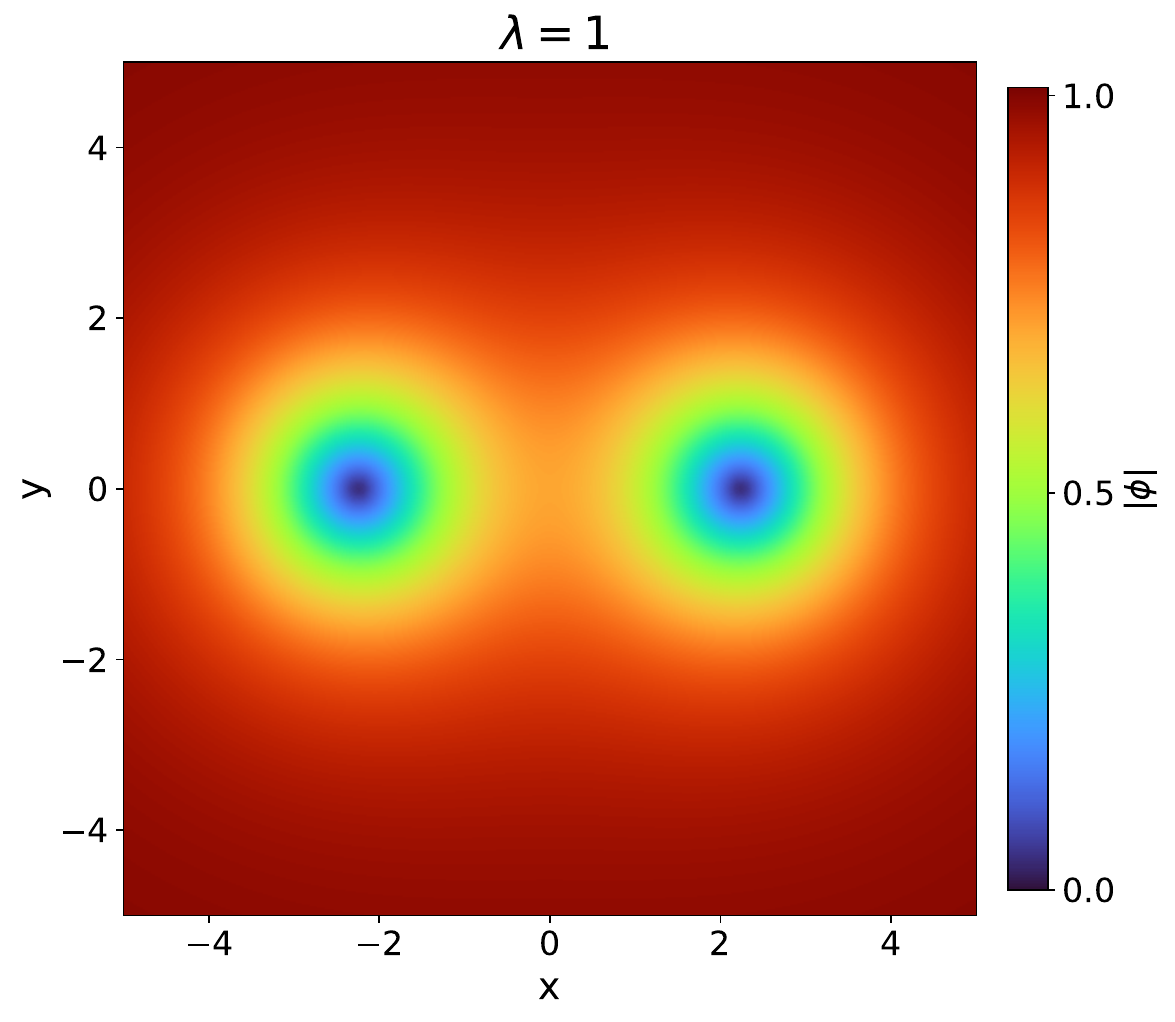}\includegraphics[width=0.4\linewidth]{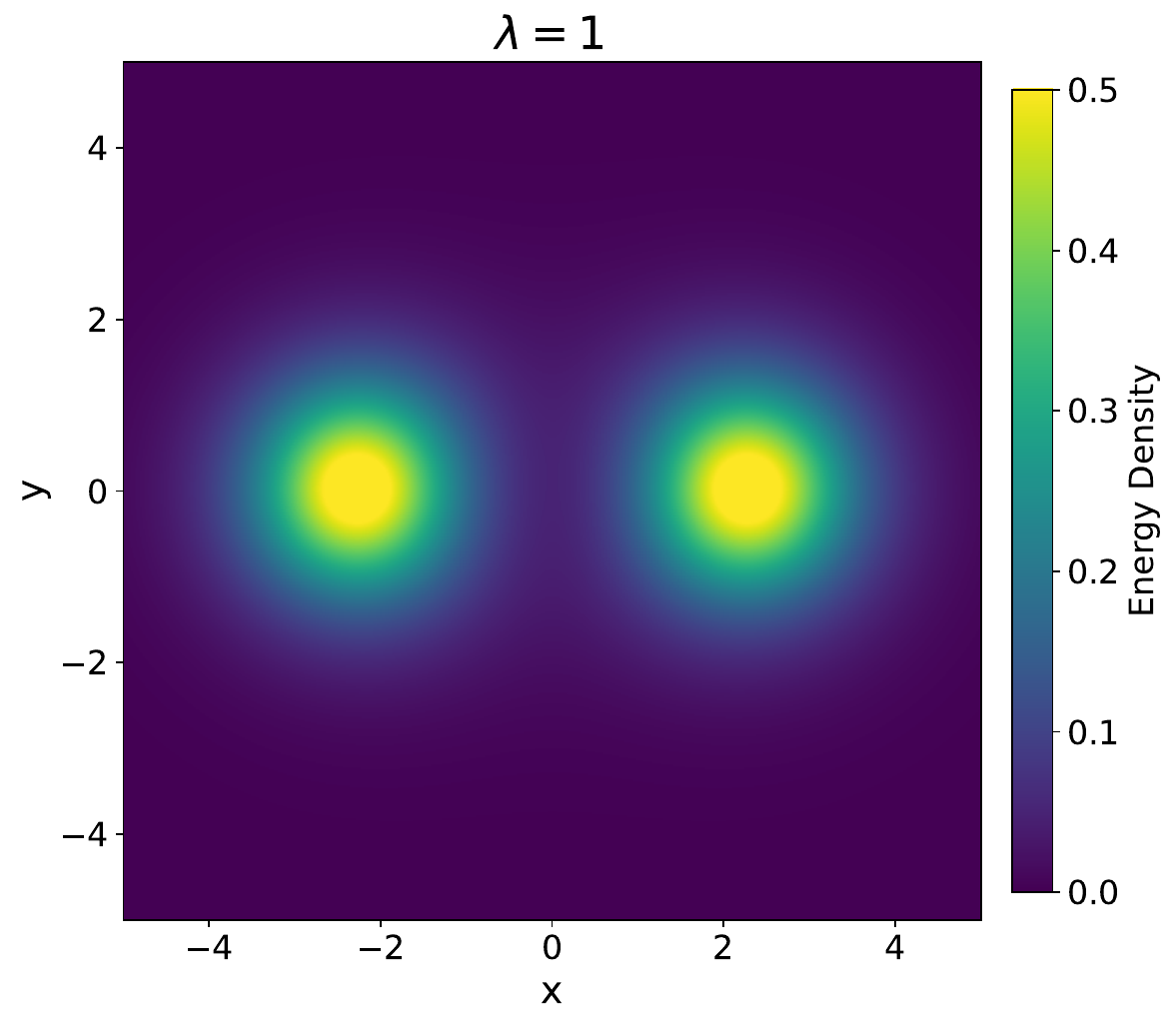}      
    \caption{\small Inner-vortex molecules for $\sigma = 2$ at $\lambda = 0.8$ and $\lambda = 1.0$. We show the absolute value of the scalar field (left) and the energy density (right).}
    \label{f:Molecule_Inner}
\end{figure}

To demonstrate the formation of a vortex molecule dynamically, we initially place two inner-vacuum vortices at a finite separation and evolve the configuration using gradient flow. We observe that the vortices move towards one another (or away from each other, depending on initial separation) until the system relaxes to a stationary configuration in which they remain separated by a finite distance $d^*(\lambda,\sigma)$. This provides a direct numerical illustration of the formation of a bound state. Fig. \ref{f:Molecule_Inner} shows the resulting configurations for $\sigma=2$ at two representative values of $\lambda$: $\lambda = 0.8$, where the energy of the $2$-vortex is lower than the energy of two well-separated $1$-vortices and $\lambda = 1.0$, where the energy is higher. The corresponding distances are in complete agreement with the interaction energies presented in Fig. \ref{fig:InteractionEnergy}. 

\begin{figure}[ht!]
    \centering
\includegraphics[width=0.6\linewidth]{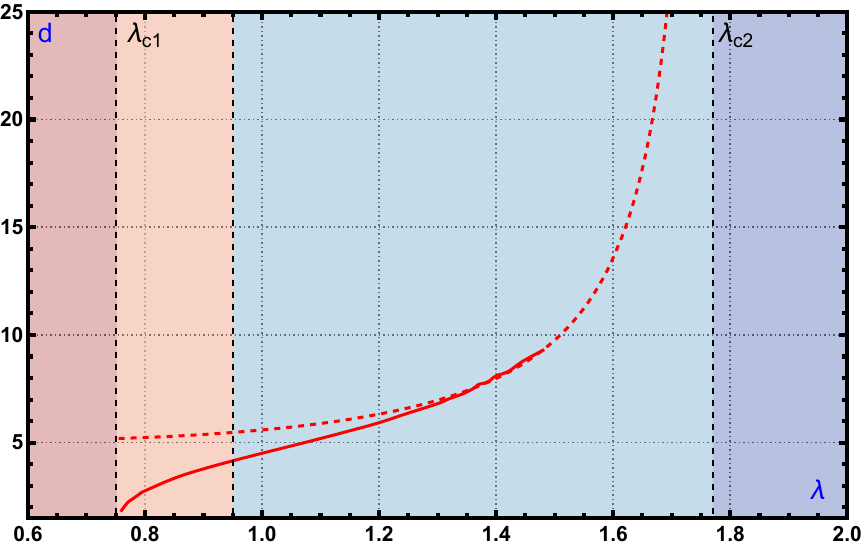}
\caption{\small 
Separation between inner $1$-vortices in the vortex molecule configuration as a function of $\lambda$ for $\sigma = 2$. Here, the dashed vertical lines represent $\lambda_{c1}$, $\lambda_{c2}$, and the value of $\lambda$ at which the energy of a $2$-vortex equals that of two well-separated $1$-vortices. The solid curve shows the separation obtained using gradient flow, while the dashed curve represents the approximate semi-analytical prediction given by \eqref{e:d_star}.}
    \label{f:dVsLambda_Inner}
\end{figure}

Fig. \ref{f:dVsLambda_Inner} shows the dependence of the vortex molecule separation on the self-coupling constant $\lambda$, as obtained using the gradient flow method. We remark that, although the interaction energy exhibits a minimum at a finite separation throughout the range $\lambda_{c1} < \lambda < \lambda_{c2}$, we are unable to numerically obtain the vortex molecule over the entire parameter range. In particular, for values of $\lambda$ close to $\lambda_{c2}$, the minimum becomes increasingly shallow, making it difficult for the gradient flow to converge to the bound configuration and causing the vortices to separate. The range of $\lambda$ for which the molecule is numerically obtained using gradient flow is indicated by the solid red curve.

Remarkably, we can estimate the equilibrium separation beyond this range by finding the extrema of the approximate interaction energy. In the large-separation limit, $d \gg 1$, the local minimum occurs at
\begin{align}\label{e:d_star}
    d^{*} \approx \frac1{m_g^{(\nu)}-m_H^{(\nu)} } \log \left[ \left(\frac{m}{q\, m_{g}^{(\nu)}}\right)^2 \sqrt{\frac{m_{g}^{(\nu)}}{m_H^{(\nu)}}}\right]\,.
\end{align}  
We show this approximation as the red dashed curve in Fig. \ref{f:dVsLambda_Inner}. Good agreement is observed for $\lambda \geq 1.25$, and the approximation appears to remain valid for larger values of $\lambda$. However, for $\lambda>1.5$ we cannot obtain the distance properly from the numerical data, since the energy minimum is too shallow to be properly measured. For $\lambda < 1.1$, the analytical agreement deteriorates because the interaction energy given by \eqref{eq:U_int} becomes less accurate at moderate and small separations, where higher-order corrections become significant. Nevertheless, this approximation remains useful as a lowest-order estimate of the separation between the inner-vacuum vortices in the molecule.

\section{Vortex-filled bubble}\label{s:Vortex_DW}

Our analysis so far has been devoted to inner-vacuum vortices. Outer vortices, however, exhibit a different structure, as illustrated in Fig. \ref{fig:Profiles_Outer}. Specifically, they can be viewed as inner-vacuum vortices surrounded by a domain wall in certain parameter regimes. This additional structure may significantly affect their interactions and dynamics, motivating a separate analysis of the outer-vortex sector.

\begin{figure}[ht!]
    \centering
    \includegraphics[width=1.02\linewidth]{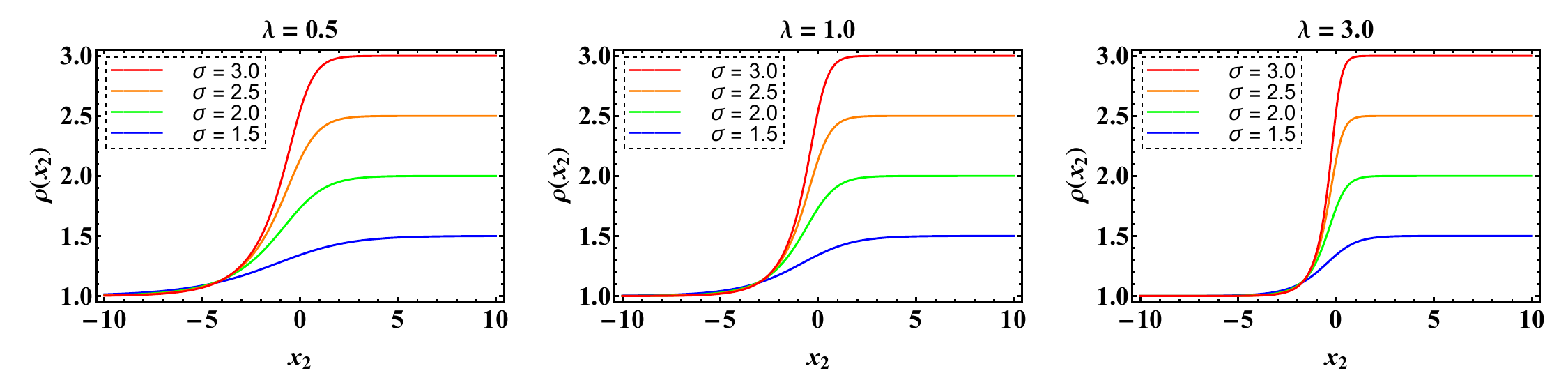}
    \caption{\small Domain wall profiles $\rho(x_2)$ for different values of $\lambda$ and $\sigma$.}
    \label{fig:DW}
\end{figure}

To begin with the analysis, let us compute the domain wall profile. 
A similar domain wall configuration interpolating two disjoint $S^1$s was discussed in Ref.~\cite{Naganuma:2000gu}.
In order to do that, let us assume a planar wall normal to the $x_2$-direction, where we assume a constant phase and, without loss of generality, choose it to vanish and set  $A_i(\vec{x},t) = 0$. Writing the scalar field profile as $\phi(x_1,x_2) = \rho(x_2)$, the wall tension, defined as the energy $V$ per unit length $L$, is given by
\begin{equation}
    T_W = \dfrac{V}{L} =  \int_{\mathbb{R}} \left[\dfrac{1}{2}\left(\dfrac{d \rho}{d x_2} \right)^2 + \dfrac{\lambda}{8}U(\rho^2)\right]\,dx_2\,.
\end{equation}
The energy density can be arranged into a Bogomolny form, from which the first-order equation for the domain-wall profile follows,
\begin{equation}
    \dfrac{d\rho}{dx_2} = \pm\dfrac{\sqrt{\lambda}}{2 \sigma^2}
    \left(\rho^2 - 1\right)
    \left(\sigma^2-\rho^2\right)\,.
\end{equation}
The boundary conditions connecting the two neighboring vacua are $\rho(-\infty) = 1$, $\rho(\infty) = \sigma$ or the configuration with the opposite orientation. The Bogomolny equation can be integrated implicitly to give
\begin{equation}\label{Eq:dw}
    \log \left(\left(\dfrac{\rho - 1}{\rho+1}\right)^{\sigma}\frac{\rho +\sigma}{\sigma -\rho }\right)= \pm \sqrt{\lambda}\frac{\sigma^2-1}{\sigma}(x_2-c) \equiv \pm m_H^{(\sigma)}(x_2-c)\,.
\end{equation}
An asymptotic analysis shows that domain walls approached the vacua exponentially. In particular, its asymptotic tails are given by
\begin{align}\label{e:asymptotics_DW}
    \rho(x_2) &\approx 1 + C_1 e^{\pm m_H^{(1)}(x_2 - c)}\,, \quad  \pm( x_2 - c)\ll - 1/m_H^{(1)}\,,\\
    \rho(x_2) &\approx \sigma - C_2 e^{\mp m_H^{(\sigma)}(x_2 - c)}\,, \quad \pm( x_2 - c) \gg 1/m_H^{(\sigma)}\,,
\end{align}
where $C_1,C_2$ are given by
\begin{equation}
     C_1= 2\left(\frac{\sigma-1}{\sigma+1}\right)^{1
 /\sigma}, \quad  C_2= 2\sigma\left(\frac{\sigma-1}{\sigma+1}\right)^\sigma.
\end{equation} 
Representative profiles of the domain wall for different
values of $\lambda$ and $\sigma$ are shown in Fig. \ref{fig:DW}.

Using the Bogomolny equation, the tension can be expressed as
\begin{align}
    T_W = \int_{-\infty}^{\infty}
    \left(\frac{d\rho}{dx_2}\right)^2 dx_2 = \frac{\sqrt{\lambda}}{2\sigma^2}
    \int_1^\sigma
    (\rho^2-1)(\sigma^2-\rho^2)d\rho = \frac{\sqrt{\lambda}}{15\sigma^2} (\sigma-1)^3
    \left(\sigma^2+3\sigma+1\right)\,.
\end{align}
Note that the tension of the wall is a monotonically increasing function where $T_W \rightarrow 0$ as $\sigma \rightarrow 1$ and $T_W \rightarrow \infty$ as $\sigma \rightarrow \infty$. If we realize the outer-vacuum vortex as a domain wall surrounding the inner-vacuum vortex, we can expect that for low values of $\sigma$, the position of the wall is far from the inner-vacuum vortex, whereas for high values the tension of the wall provokes that the wall collapses toward the origin, completely merging with the inner-vacuum vortex. This is indeed seen in the energy density profiles in Fig. \ref{fig:Energy_Magnetic_Outer}. 

\begin{figure}[ht!]
    \centering
    \includegraphics[width=0.95\linewidth]{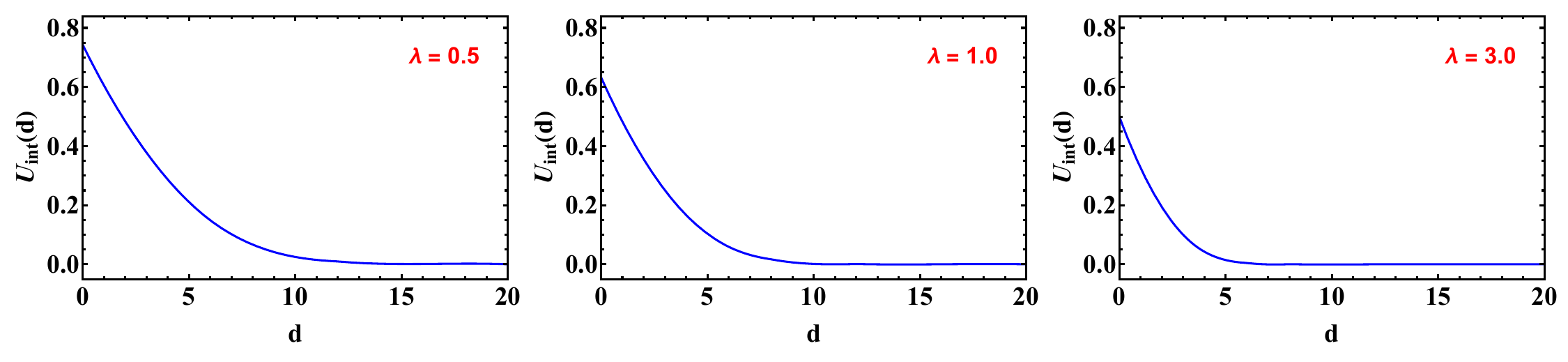}
    \includegraphics[width=0.95\linewidth]{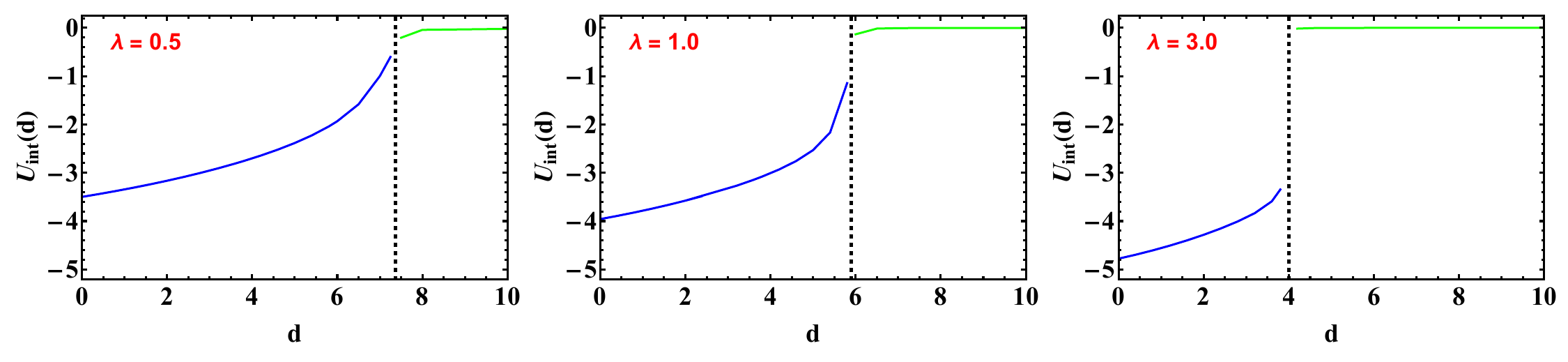}
    \caption{\small Interaction energy between the domain wall and the inner-vacuum vortex (upper panel) or the outer-vacuum vortex (lower panel) for $\sigma = 2.0$.}
    \label{fig:DW_Vortex}
\end{figure}

Having obtained the domain wall profile, we can now analyze the interaction energy between the domain wall and the different species of vortices. To this end, we once again employ a relaxation method, taking the following ansatz as the initial configuration
\begin{eqnarray}
    \phi^{DW-V}(\vec{x};d) &=& \dfrac{1}{v}\rho(x_2)\phi^{(1)}(\vec{x};d)\,, \\ 
    A_i^{DW-V}(\vec{x};d) &=& A_i^{(1)}(\vec{x};d)\,,
\end{eqnarray}
which describes a domain wall $\rho(x_2)$ extending along the $x$ direction and a vortex $(\phi^{(1)},A_i^{(1)})$ located at $(x_1,x_2) = (0,\pm d)$. The parameter $d$ therefore specifies the initial separation between the vortex centre and the domain wall. The positive and negative signs correspond to outer and inner-vacuum vortices, respectively, given our choice of domain-wall orientation. During the relaxation, the vortex is pinned at $(x_1,x_2) = (0,\pm d)$, while the domain wall is anchored at $(x_1,x_2) = (\pm 0.95 L,0)$. These constraints on the domain wall prevent an overall displacement of the wall during the relaxation, and should be assumed sufficiently far away from the vortex so as not to affect its dynamics. The interaction energies between the domain wall and the inner and outer-vacuum vortices, defined as $U_{int}(d) = V^{DW-V}(d) - V[\rho] - V[\phi^{(1)},A^{(1)}]$, are shown in the upper and lower panels of Fig. \ref{fig:DW_Vortex}, respectively, for different values of $\lambda$ at $\sigma = 2$. Here, $V^{DW-V}(d)$ denotes the relaxed configuration at a distance $d$ between the domain wall and a vortex, $V[\rho]$ is the static energy of the domain wall, and $V[\phi^{(1)},A^{(1)}]$ is the static energy of the vortex. Once more, we have performed the simulations in a box of size $[-L,L]\times[-L,L]$ with $L = 30$.

Our results reveals that the inner-vacuum vortex is repelled by the domain wall whilst the outer-vacuum vortex is attracted. We have corroborated this through two different procedures. On the one hand, we have followed the approach developed in Ref. \cite{Manton1997} to determine the interaction energy between kinks, i.e., we calculate the force between the vortex and the domain wall by integrating the appropriate component of the energy-momentum tensor a line separating the vortex and the domain wall. Using the asymptotic field profiles \eqref{e:asymptotic} and \eqref{e:asymptotics_DW}, we obtain the force at large separations and then integrate it to determine the interaction energy. Another possibility consists in adapting the method introduced in Ref. \cite{Speight1997} to determine the interaction between vortices.  In both cases, we find that, at leading order, the domain wall and the vortices only interact through the scalar channel, yielding the following interaction energies
\begin{equation}\label{Eq: IntEnDw}
U_{\text{int}}^{(1)}(d)=2\,q^{(1)}\,\left(\frac{\sigma-1}{\sigma+1}\right)^{1/\sigma} e^{-m^{(1)}_H d}\,,\quad U_{\text{int}}^{(\sigma)}(d)=-2\,\sigma \,q^{(\sigma)}\,\left(\frac{\sigma-1}{\sigma+1}\right)^\sigma e^{-m^{(\sigma)}_H d} \,,
\end{equation}
where $d$ is the distance between the vortex and the domain wall.
Therefore, our approximated analytical expressions \eqref{Eq: IntEnDw} predict the numerical results presented in Fig. \ref{fig:DW_Vortex}. 

\begin{figure}[ht!]
        \centering
        \includegraphics[width=0.45\linewidth]{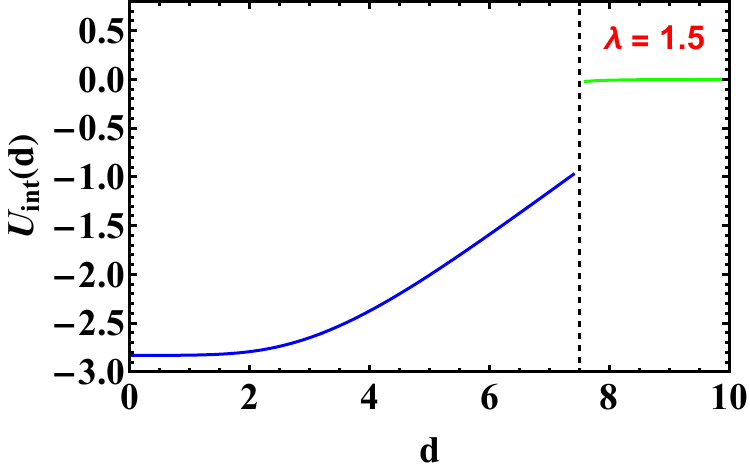}
        \includegraphics[width=0.45\linewidth]{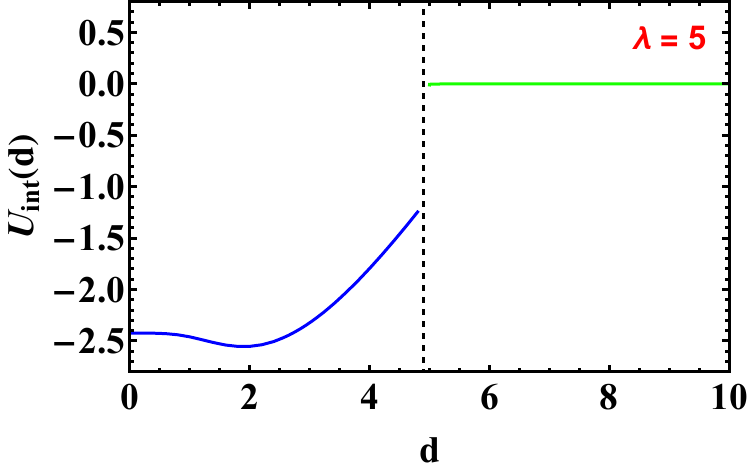}%
          
        \caption{\small Vortex-filled bubble for $\sigma = 2$ and $\lambda = 1.5$ (upper panel) and $\lambda = 5$ (lower panel). The dashed line denotes the distance where we observe the phase transition.} 
    \label{f:InteractionEnergy_Outer}
\end{figure}

With this insight we can now tackle the problem of interaction between outer-vacuum vortices. The relaxation method is formally identical to the case of inner-vacuum vortices, so we consider the initial ansatz \eqref{e:Abrikosov_Vortices} with $v = \sigma$. We display the interaction energy between outer-vacuum vortices in Fig.~\ref{f:InteractionEnergy_Outer} for different values of $\lambda$ at $\sigma = 2$. Through this analysis we observe that the outer-vacuum vortices also exhibit a nonmonotonic interaction energy allowing for the formation of stable intermediate states. Following the same reasoning as for inner-vacuum vortices, we would expect a local maximum in the interaction energy slightly above the critical value $\gamma=1$, where $q\gg m/\sigma$ (see Fig. \ref{fig:q_m} lowe panel). However, this feature is masked by the presence of a phase transition, that we denote by the dashed line in Fig.~\ref{f:InteractionEnergy_Outer}.

\begin{figure}[ht!]
    \centering

        \includegraphics[width=0.4\linewidth]{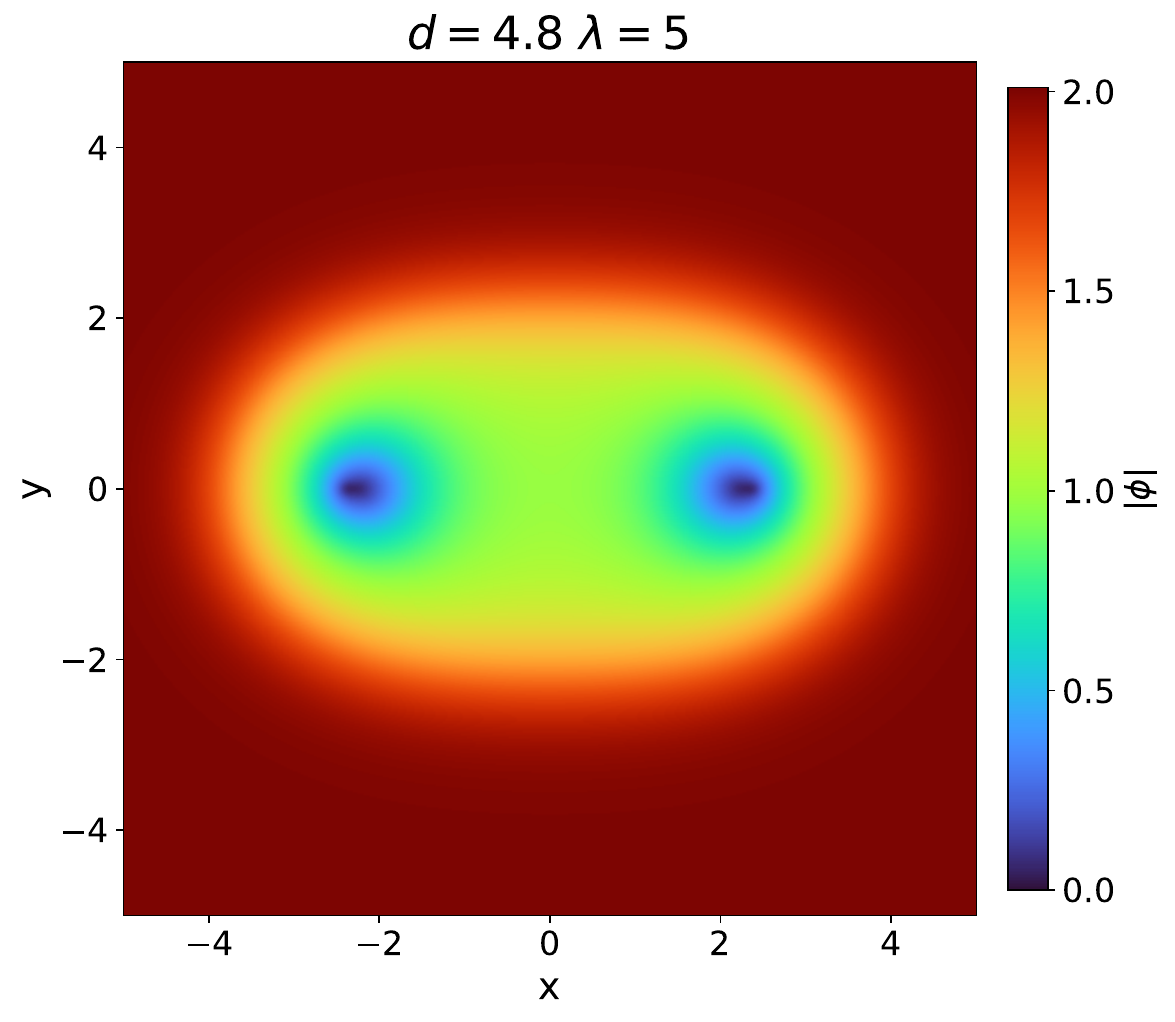}\includegraphics[width=0.4\linewidth]{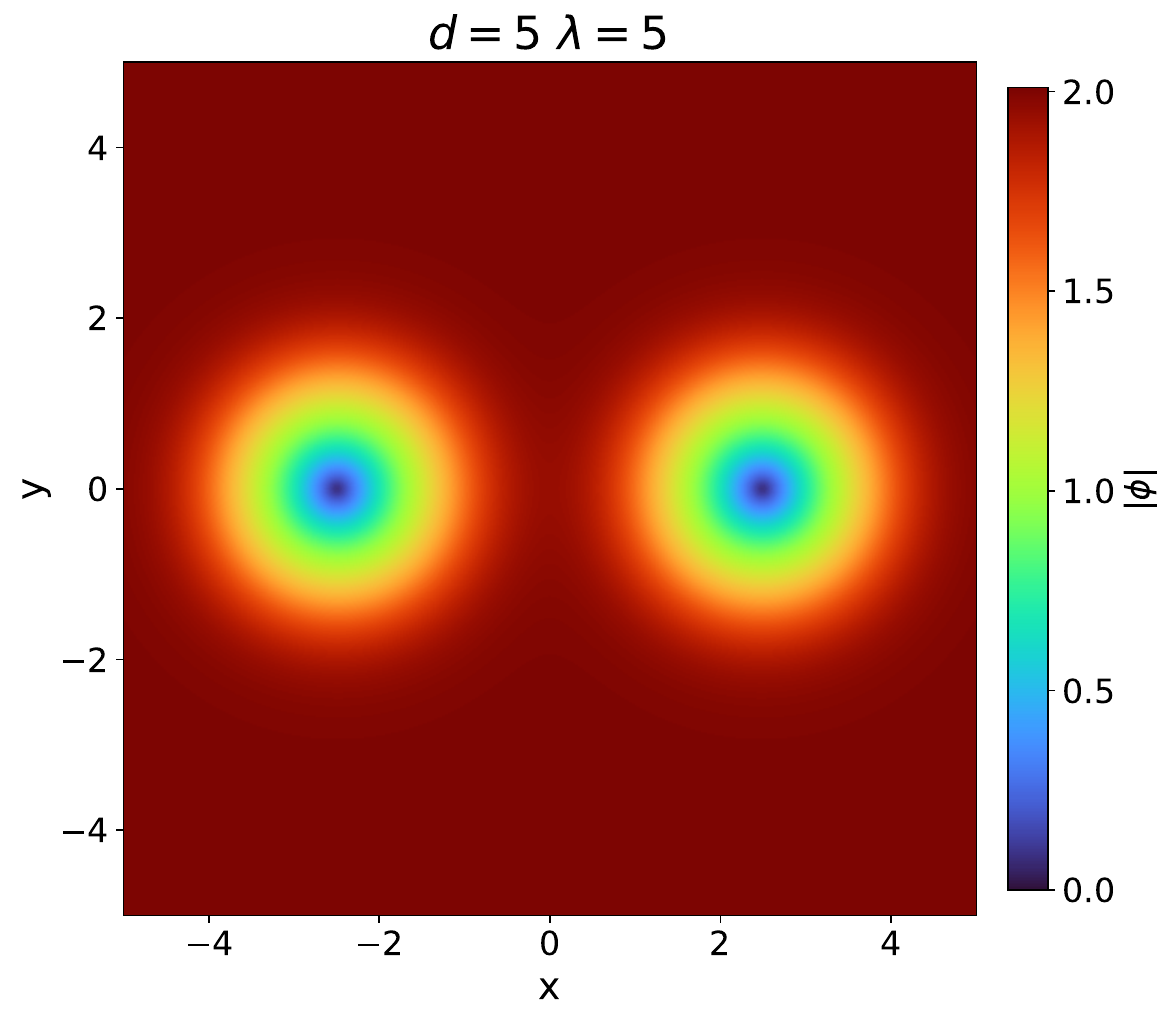}%
        \caption{\small Transition between the two different mimimal energy configurations between two near outer-vacuum vortices for $\lambda = 5$ and $\sigma = 2$.}
    \label{f:PhaseTransition}
\end{figure}

When the two outer-vacuum vortices are far apart, the interaction energy is governed strictly by the overlap of their asymptotic tails. However, since the Higgs mass $m_H^{(\sigma)}$ and the gauge mass $m_g^{(\sigma)}$ are higher than for the inner-vacuum vortices, they only interact when they are sufficiently close. As the separation decreases, the system reaches a threshold where the individual two-vortex state is no longer the energy-minimizing configuration. At this stage, individual inner-vacuum vortices become enveloped by a domain-wall-like structure. This configuration resembles a trapped multi-vortex state confined within a domain boundary rather than two smoothly merging individual vortices. This behavior is clearly confirmed in Fig. \ref{f:PhaseTransition}, where the modulus of the relaxed scalar field $|\phi|$ is shown at a distance before and after the jump for $\sigma = 2$. 

\begin{figure}[ht!]
        \centering
        \includegraphics[width=0.4\linewidth]{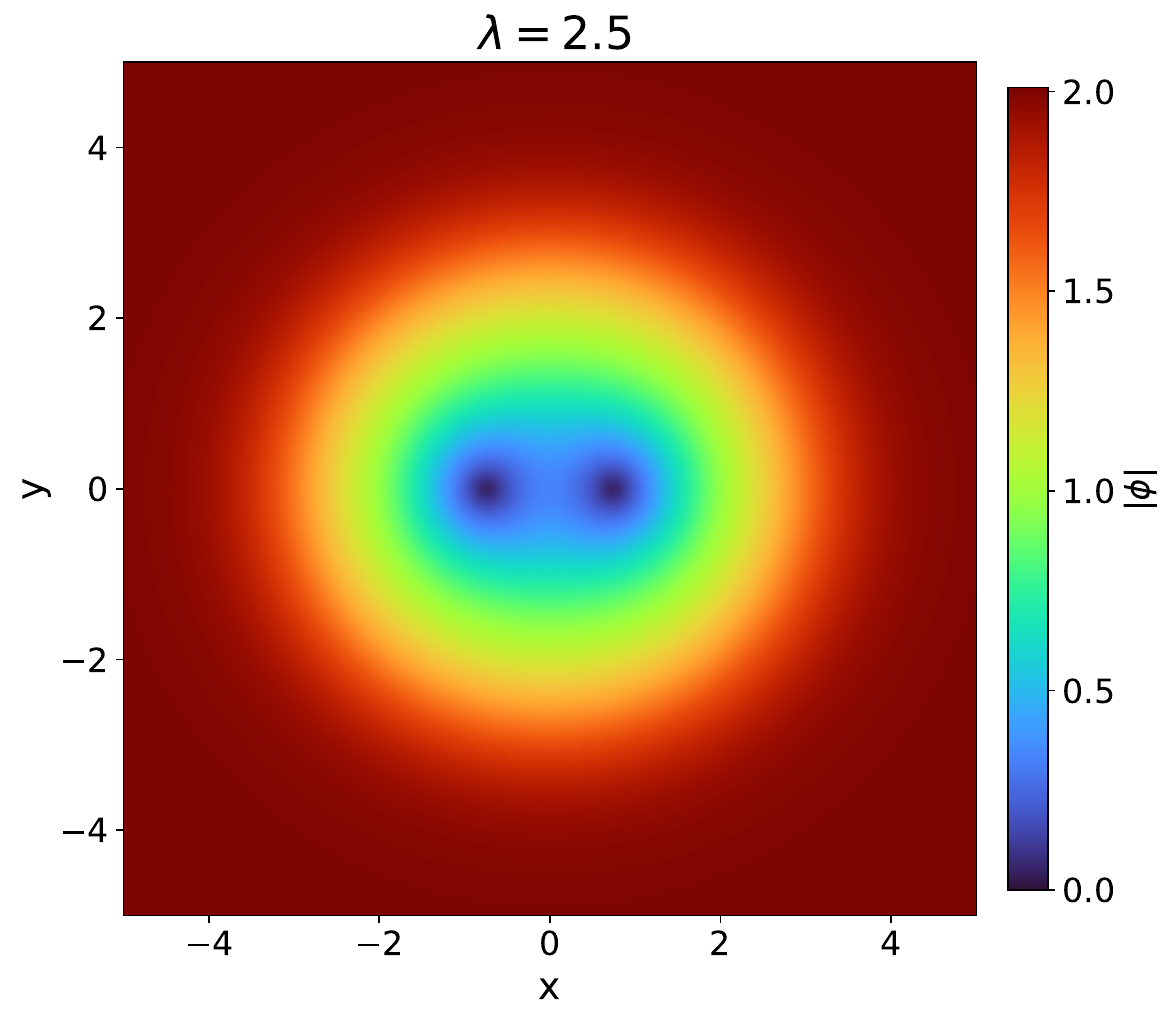}\includegraphics[width=0.4\linewidth]{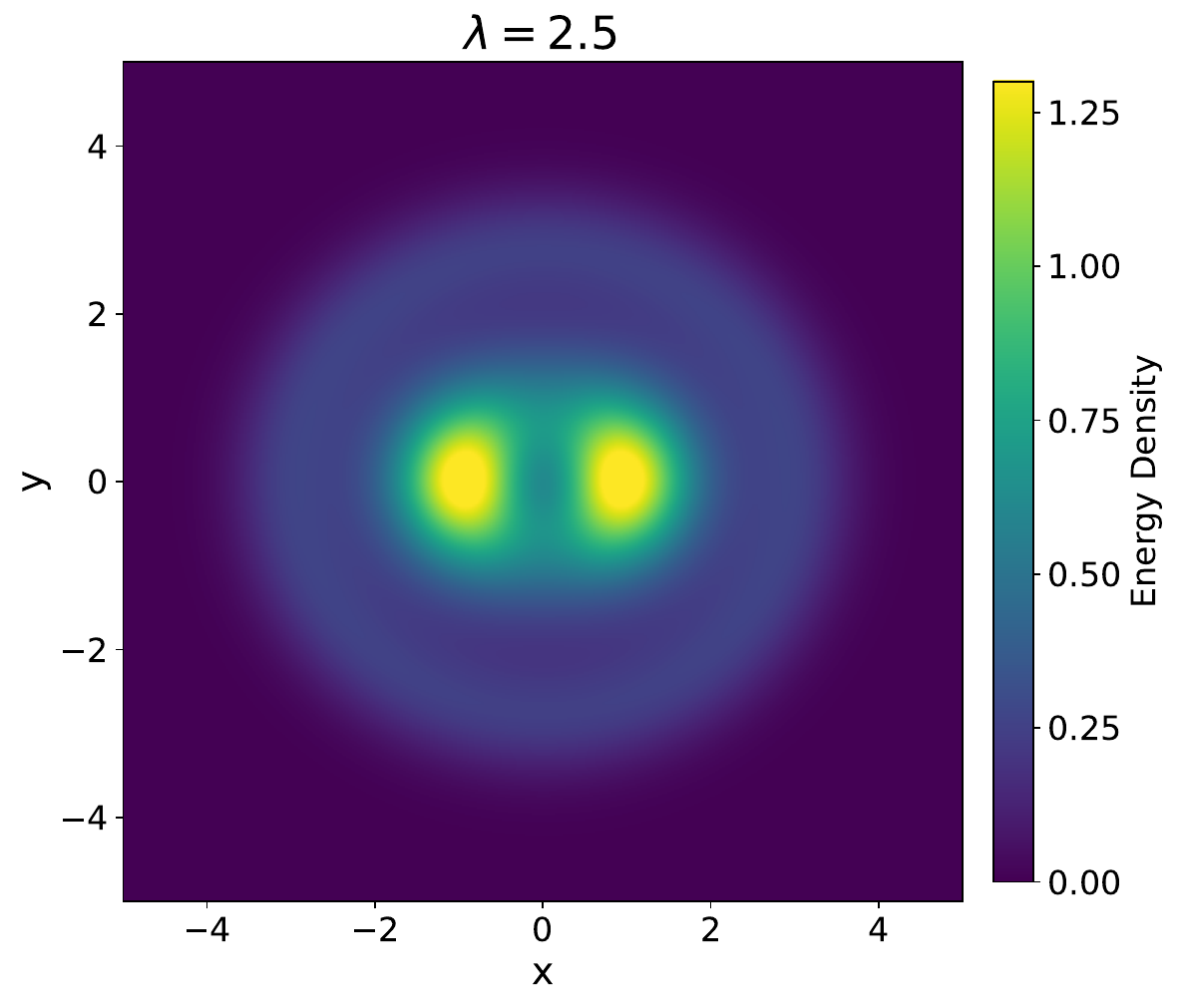}%
        
        \centering
        \includegraphics[width=0.4\linewidth]{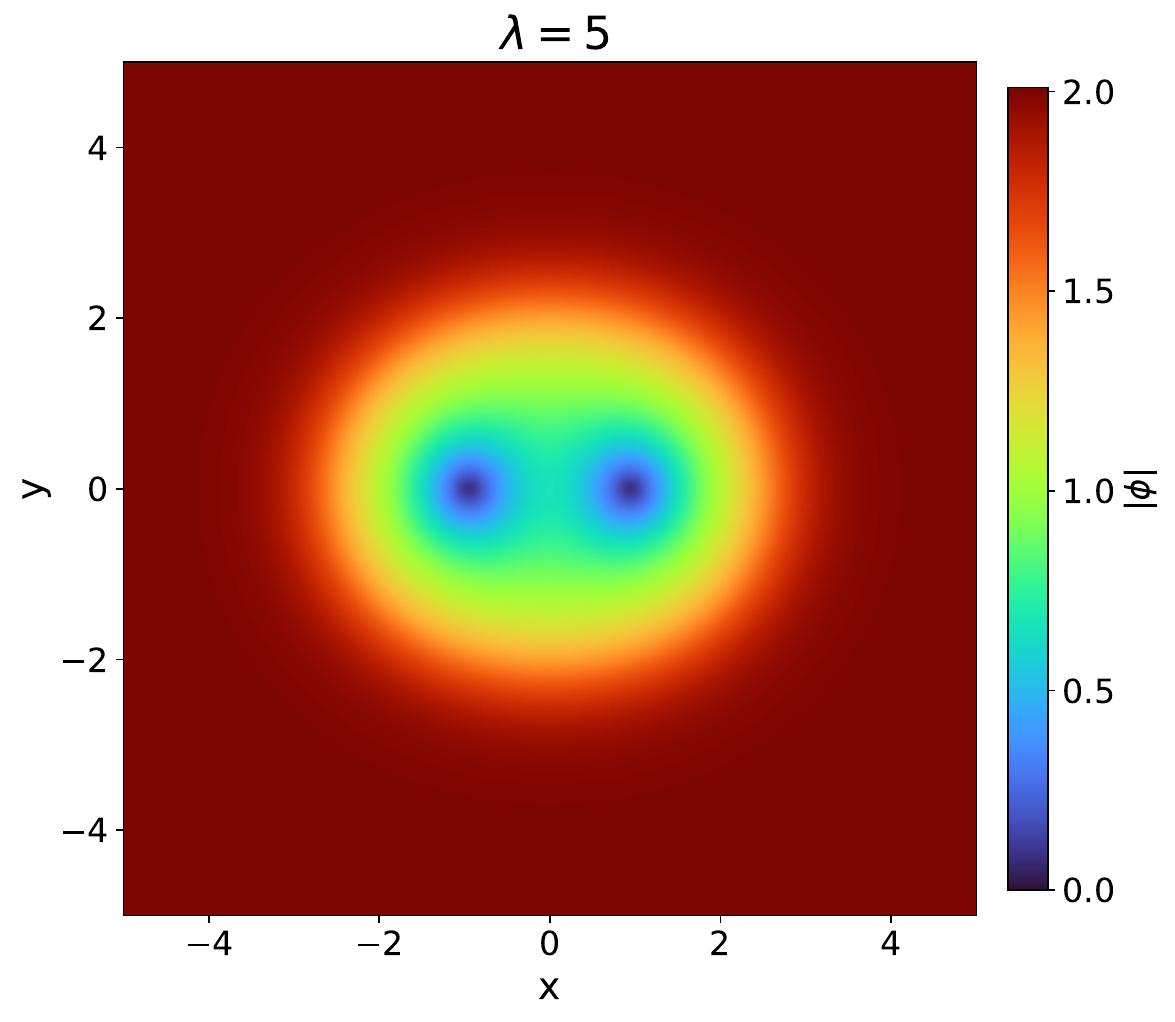}\includegraphics[width=0.4\linewidth]{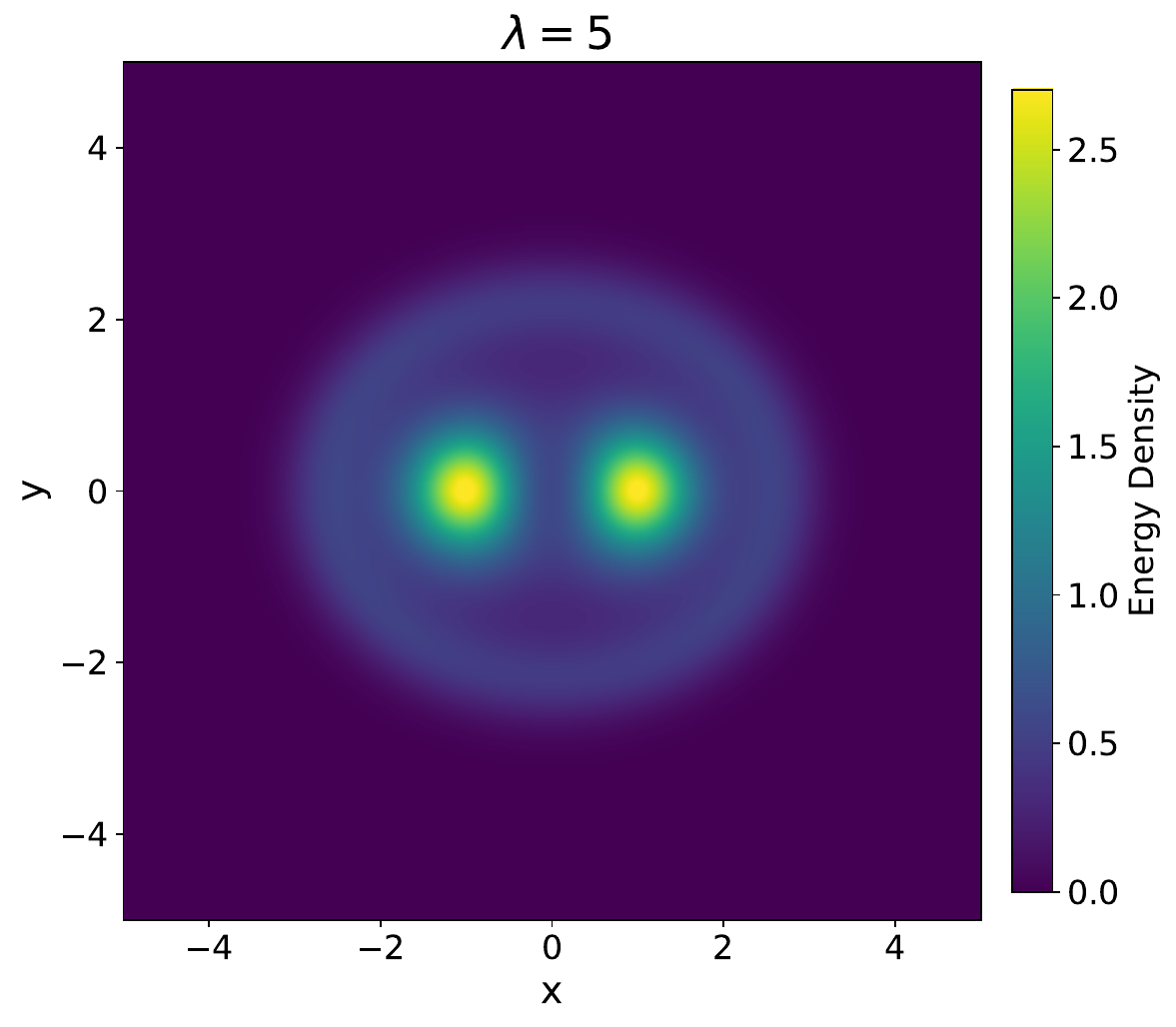}%
          
        \caption{\small Vortex-filled bubble for $\sigma = 2$ and $\lambda = 2.5$ (upper panel) and $\lambda = 5$ (lower panel). We represent the modulus of the scalar field and the energy density.}
    \label{f:Molecule_Outer}
\end{figure}

\begin{figure}[ht!]
    \centering
\includegraphics[width=0.6\linewidth]{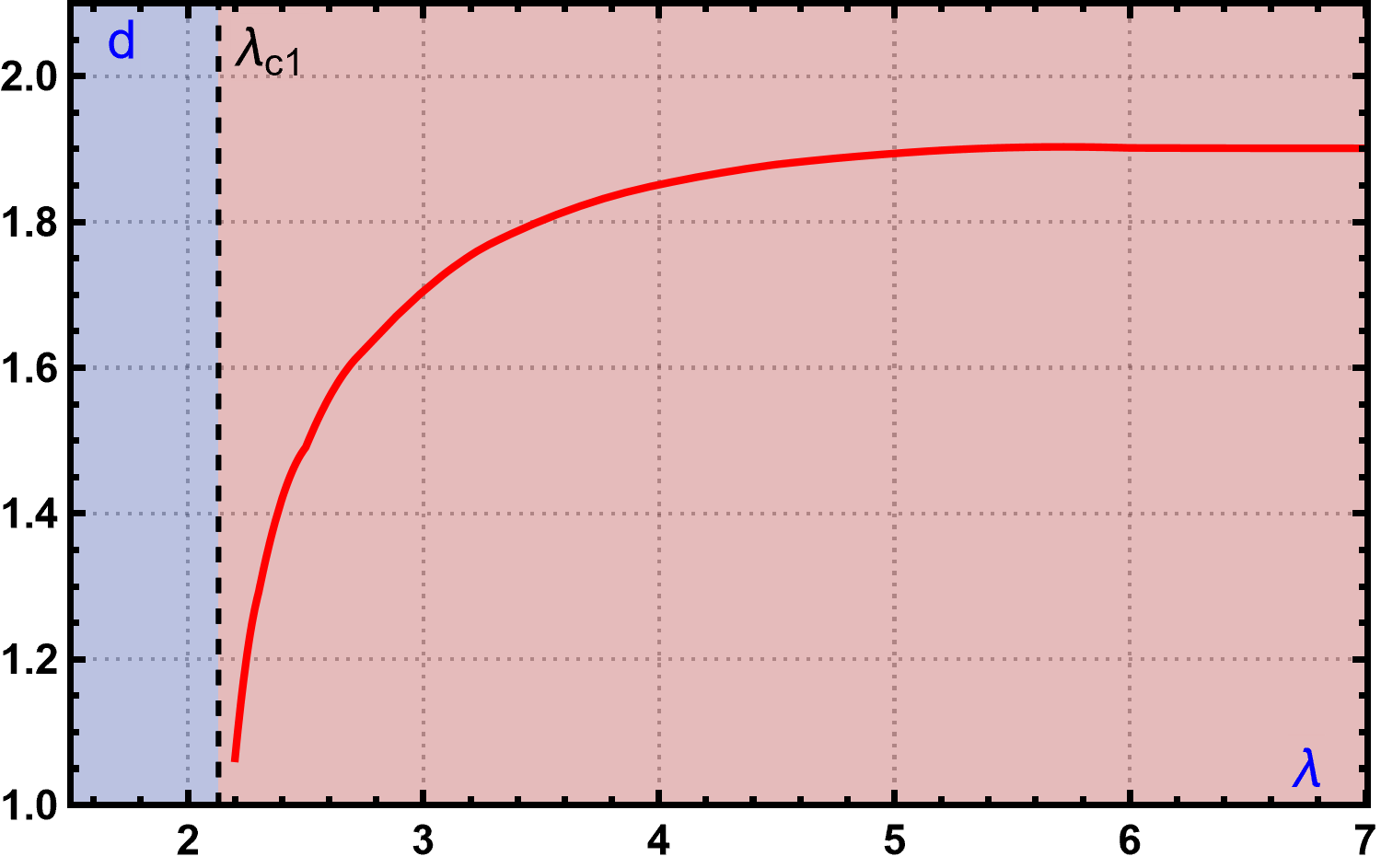}    \caption{\small 
Separation between outer $1$-vortices in the vortex molecule configuration as a function of $\lambda$ for $\sigma = 2$. Note that $\lambda_{c1} \approx 2.15$ is higher than $\lambda_{c2} \approx 1.78$.}
    \label{f:dVsLambda_Outer}
\end{figure}

As expected, the interaction between outer-vacuum vortices differs fundamentally from that between inner-vacuum vortices. In the regime where an outer-vacuum vortex can be viewed as an inner-vacuum vortex surrounded by a domain wall, two approaching outer-vacuum vortices can be described in terms of their constituent inner-vacuum vortices confined within a common bubble. In this picture, the equilibrium is governed by the competition between the inward repulsive force exerted by the surrounding domain wall and the outward repulsive vector interaction between the inner-vacuum vortices. Two examples of stable vortex-fillex bubbles are displayed in Fig. \ref{f:Molecule_Outer}. Unlike the case of isolated inner-vacuum vortices, this mechanism suggests that stable configurations may persist for arbitrarily large values of $\lambda$, with no upper critical value. Indeed, we expect the lower critical value $\lambda_{c1}$ for the existence of a nontrivial equilibrium to satisfy $\lambda_{c1} > \lambda_{c2}$. The reason is that, within this picture, the two inner-vacuum vortices would otherwise collapse into a single doubly quantized vortex unless the repulsive vector interaction is sufficiently strong to overcome their tendency to collapse. Fig. \ref{f:dVsLambda_Outer} displays the position of the global minimum with respect to the self-coupling constant $\lambda$. As explained, the minimum appears at a value of the parameter $\lambda$ higher than that for which the masses are equal, specifically, $\lambda_{c1} \approx 2.15$ for $\sigma = 2$. Surprisingly, the separation between inner-vacuum vortices in the vortex-filled bubble tends to a constant distance at high values of $\lambda$. This behavior contrasts with that observed for inner-vacuum vortices, for which the equilibrium separation increases with $\lambda$.   

\section{Spectrum of perturbations around the vortex} \label{s:LinearProblem}

The previous analysis of the interaction energy between vortices suggests that coincident $n$-vortex configurations may be either stable or unstable, depending on the regime under consideration. This distinction is encoded in the linearized spectrum, where instability is reflected in the presence of unstable modes \cite{GoodbandHindmarsh1995}. We therefore turn to the spectral analysis of fluctuations around the coincident vortices. This study provides valuable insight into their stability and also their dynamical properties.

Let us call the scalar and vector components of the axially-symmetric static $n$-vortex \eqref{e:ANO} as
\begin{equation}
    \Psi(\vec{x};n) = \begin{pmatrix}
    \psi_1(\vec{x};n)\\
    \psi_2(\vec{x};n)\\
    V_1(\vec{x};n)\\
    V_2(\vec{x};n)
    \end{pmatrix} 
    = \begin{pmatrix}
    f_n(r)\cos(n \,\theta)\\
    f_n(r)\sin(n\, \theta)\\
    - \dfrac{n\beta_n(r)}{r}\sin \theta\\
    \dfrac{n\beta_n(r)}{r}\cos \theta
    \end{pmatrix}\,.
\end{equation}
An infinitesimal perturbation about the $n$-vortex configuration can then be written as
\begin{equation}\label{e:Perturbation}
    \widetilde{\Psi}(\vec{x};n) = \Psi(\vec{x};n) +  \epsilon\,\xi(\vec{x};n)\,,
\end{equation}
where
\begin{equation}
    \xi(\vec{x};n) = \begin{pmatrix}
    \varphi_1(\vec{x};n)\\
    \varphi_2(\vec{x};n)\\
    a_1(\vec{x};n)\\
    a_2(\vec{x};n)
    \end{pmatrix}\,,
\end{equation}
and $\epsilon$ is the amplitude of the perturbation. To discard the unphysical gauge degrees of freedom, we impose the background gauge condition
\begin{equation}
    \partial_1a_1(\vec{x}) + \partial_2a_2(\vec{x}) - \left(\psi_1(\vec{x})\varphi_2(\vec{x}) - \psi_2(\vec{x})\varphi_1(\vec{x}) \right) = 0\,.
\end{equation}
This condition fixes the residual gauge freedom in the perturbation modes. Substituting the perturbed configuration \eqref{e:Perturbation} into the equations of motion \eqref{Eq:EquationSecondOrdenCartesianCoordinates} and retaining terms to first order in $\epsilon$, we obtain the following eigenvalue problem
\begin{equation}
    {\cal H}^+\xi_\nu(\vec{x};n) = \omega^2_{\nu}\,\xi_\nu(\vec{x};n)\,,
\end{equation}
where
    \begin{equation}\label{eq:SecondOrderH}
    \hspace{-1cm}
	{\cal H}^+= \left( \begin{array}{cccc}
	-\Delta + F(\psi_1,\psi_2)  +V_kV_k & -2 V_k \partial_k + C(\psi_1,\psi_2) & -2 \widetilde{D}_1 \psi_2 & -2 \widetilde{D}_2\psi_2 \\
	2V_k \partial_k + C(\psi_1,\psi_2) & -\Delta + F(\psi_2,\psi_1) + V_kV_k & 2\widetilde{D}_1\psi_1 & 2 \widetilde{D}_2 \psi_1 \\
	-2 \widetilde{D}_1 \psi_2 & 2 \widetilde{D}_1 \psi_1 & -\Delta + |\psi|^2 & 0 \\
	-2 \widetilde{D}_2 \psi_2 & 2 \widetilde{D}_2 \psi_1 & 0 & -\Delta +|\psi|^2  
	\end{array} \right)
\,,
\end{equation}
with $\widetilde{D}_i\,\psi_j = \partial_i\,\psi_j + \epsilon^{jk}V_i\psi_k$ and $i,j,k = \{1,2\}$\,. Here, $C(\psi_1,\psi_2)$ and $F(\psi_i,\psi_j)$ are given by
\begin{align}
C(\psi_1,\psi_2)&= \dfrac{\lambda}{\sigma^4}\psi_1\psi_2
\left(1 + 4 \sigma^2 + \sigma^4 + 6 \psi_k^2\left(\psi_k^2-\sigma^2-1\right) + 12 \psi_1^2\psi_2^2\right)\,,
\\
F(\psi_i,\psi_j) &= \dfrac{\lambda}{2\sigma^4}\left[
6\psi_i^2\psi_j^2\left(5\psi_i^2+3\psi_j^2-3\sigma^2-3\right)+\psi_i^2\left(-15\sigma^2\psi_i^2+14\psi_i^4-15\psi_i^2+3\sigma^4+12\sigma^2+3\right)
\right. \nonumber\\
&\left.\qquad +\psi_j^2\left(-3\sigma^2\psi_j^2+2\psi_j^4-3\psi_j^2+\sigma^4+4\sigma^2+1\right)
\right]\,.
\end{align}
It can be verified that, in the limit $\sigma \to \infty$, the second-order fluctuation operator \eqref{eq:SecondOrderH} reduces to the corresponding operator of the standard $\phi^4$ model \cite{GoodbandHindmarsh1995}. 

Following the same strategy as in Refs. \cite{AlonsoIzquierdoMiguelezCaballero2024,AlonsoIzquierdo2026b}, we find the same angular ansatz for the internal modes    
\begin{equation}\label{modooos}
    \xi_\nu(\vec{x};n,\overline{k}) = \left( \begin{array}{c} 
    \cos(\overline{k}\,\theta)\cos(n\,\theta) u(r) + \overline{k} \sin(\overline{k}\,\theta) \sin(n\,\theta) w(r) \\
    \cos(\overline{k}\,\theta)\sin(n\,\theta) u(r) - \overline{k} \sin(\overline{k}\,\theta) \cos(n\,\theta) w(r)\\
    \cos(\overline{k} \,\theta) \sin\theta [v'(r) -r\, f_n(r) w(r)] - \sin (\overline{k}\,\theta) \cos \theta \frac{\overline{k} v(r)}{r}  \\
    - \cos(\overline{k}\, \theta) \cos\theta [v'(r) -r\, f_n(r) w(r)] - \sin (\overline{k}\,\theta) \sin \theta \frac{\overline{k} v(r)}{r} \\
     \end{array} \right)\,,  
\end{equation}
where $\overline{k}=0,1,2,3\dots$. Here, $\bar{k} = 0$ corresponds to axially-symmetric modes, whereas the $\bar{k} \neq 0$ modes correspond to the so-called multipolar modes. As explained in Ref. \cite{AlonsoIzquierdo2026b}, the modes with $\bar{k} > 0$ are doubly degenerate, since we can find an alternative linearly independent set of modes with the same eigenfrequencies \footnote{It can be proved that the exact form for this degenerate mode can be obtained by making the changes $\cos(\overline{k}\,\theta)\rightarrow\sin(\overline{k}\theta)$, $\sin(\overline{k}\,\theta)\rightarrow-\cos(\overline{k}\theta)$ in \eqref{modooos}.}.

In our model under study, the $u(r), v(r)$ and $w(r)$ functions satisfy the following system of coupled ordinary differential equations
\begin{equation}
\begin{aligned}
-u''(r) &-\frac{u'(r)}{r}
    +\Bigg[\frac{\overline{k}^2}{r^2}
    +\frac{(n-\beta_n(r))^2}{r^2}
    -W_u(r)\Bigg]u(r) \\
    &\quad -\frac{2(n-\beta_n(r))(\overline{k}^2+r^2f_n(r)^2)}
    {r^2}w(r)
    +\frac{2f_n(r)(n-\beta_n(r))}{r}v'(r)
    =\omega_{\nu}^2u(r), \\[2mm]
    -v''(r) &-\frac{v'(r)}{r}
    +\Bigg[\frac{\overline{k}^2}{r^2}+f_n(r)^2\Bigg]v(r)
    +2w(r)\left(f_n(r)+rf_n'(r)\right)
    =\omega_{\nu}^2v(r), \\[2mm]
    -w''(r) &-\frac{w'(r)}{r}
    +\Bigg[\frac{\overline{k}^2}{r^2}
    +\frac{(n-\beta_n(r))^2}{r^2}
    +f_n(r)^2-W_w(r)\Bigg]w(r) \\
    &\quad -\frac{2(n-\beta_n(r))}{r^2}u(r)
    +\frac{2f_n'(r)}{r}v(r)
    =\omega_{\nu}^2w(r).
\end{aligned}
\end{equation}
where
\begin{align}
    W_u(r) &= \frac{\lambda }{2 \sigma ^4}\left[\sigma ^4+\sigma ^2 -3 \left(\sigma ^4+4 \sigma ^2+1\right) f(r)^2 + \dfrac{15}{2} \left(\sigma ^2+1\right) f(r)^4-14 f(r)^6\right]\,,\\
    W_w(r) &= \frac{\lambda}{2 \sigma ^4}\left(1- f(r)^2\right) \left[\sigma ^4 + \sigma ^2  + 2 f(r)^4 -\left(3 \sigma ^2+1\right) f(r)^2\right]\,.
\end{align}
To solve this spectral problem, we discretize the ordinary differential equations using a second-order finite-difference scheme. The radial coordinate is discretized as $r_i = i\Delta r$, with a grid spacing $\Delta r = L/N$, where $N = 2000$. Moreover, we impose the regularity conditions $u'(0) = v'(0) = w'(0) = 0$ as well as the boundary conditions $u(N) = v(N) = w(N) = 0$, as we are interested in the bounded modes. The resulting discretized Hessian defines a matrix eigenvalue problem, which we solve numerically using the implicitly restarted Arnoldi method.  

\begin{figure}[ht!]
    \centering

    \includegraphics[width=0.458\linewidth]{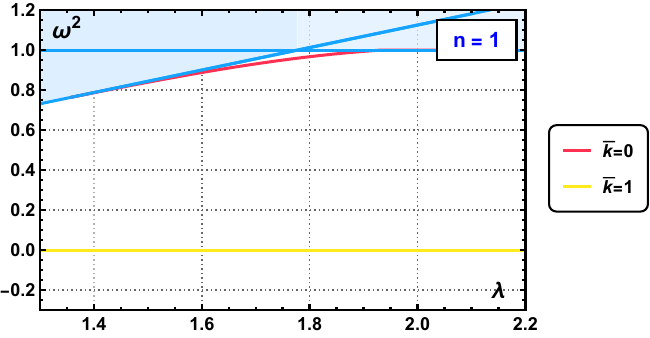}
    \hfill
    \includegraphics[width=0.45\linewidth]{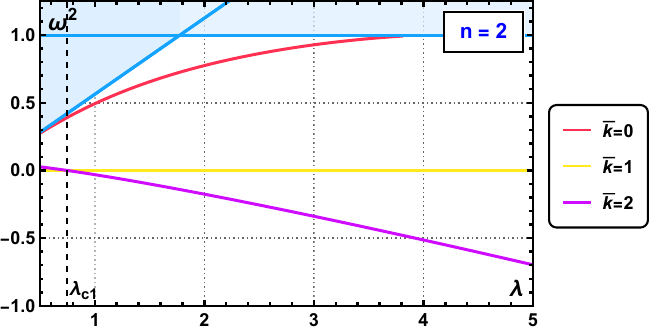}

    \vspace{0.2cm}

    \includegraphics[width=0.45\linewidth]{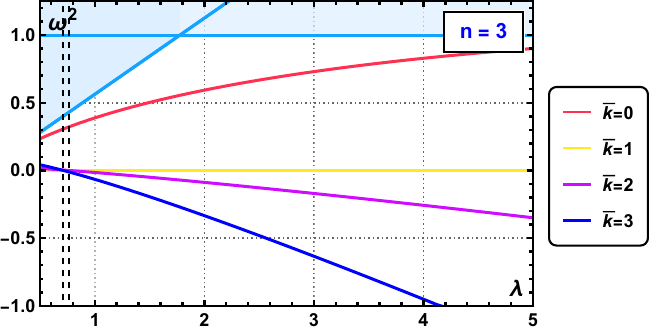}
    \hfill
    \includegraphics[width=0.45\linewidth]{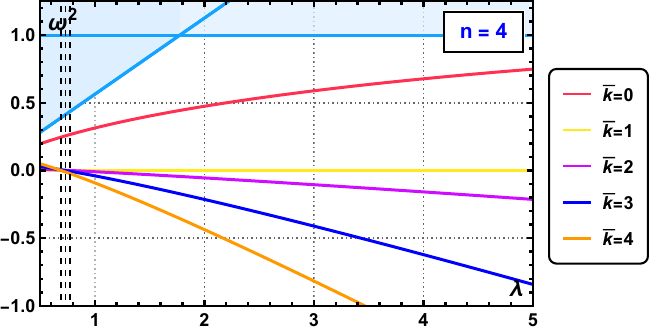}

    \caption{\small Spectral structure for the inner $n$-coincident vortex for $n=1,\ldots,4$. We have assumed $\sigma = 2$ as an illustrative example. The vertical dashed lines indicate the points at which the eigenvalue of the $2 \leq \bar{k} \leq n$ modes cross zero.}
    \label{fig:Modes}
\end{figure}

We depict the spectral structure of inner-vacuum vortices with winding number $n=1,\ldots,4$ for $\sigma = 2$ in Fig. \ref{fig:Modes}. As expected, the $n = 1$ vortex is stable, since there are no negative eigenmodes in its spectrum. It possesses a zero mode associated with translational invariance in each spatial direction, with $\bar{k} = 1$, as well as an axially symmetric Derrick-type mode with $\bar{k} = 0$. On the other hand, vortices with $n > 1$ possess modes with $2 \leq \bar{k} \leq n$, coined Type-A modes, that are positive  within a certain range of values of $\lambda$ and render negative above a certain value. Therefore, the vortex becomes unstable in the latter case. Unlike the standard Abelian-Higgs model, where all unstable modes cross the zero-frequency axis simultaneously at the BPS limit ($\lambda = 1$), in our $\phi^8$ framework each mode crosses it at a distinct value of the self-coupling constant. The values of $\lambda$ at which the multipolar modes become unstable are indicated by the dashed lines. Another difference with respect to the $\phi^4$ model is that, within the parameter range investigated, we do not observe any Type-B multipolar modes, i.e., multipolar modes whose eigenvalues satisfy $\omega^2 >0$, for all $\lambda \in \mathbb{R}$ \cite{AlonsoIzquierdo2026b}.

\begin{figure}[ht!]
    \centering
        \centering
        \includegraphics[width=0.376\linewidth]{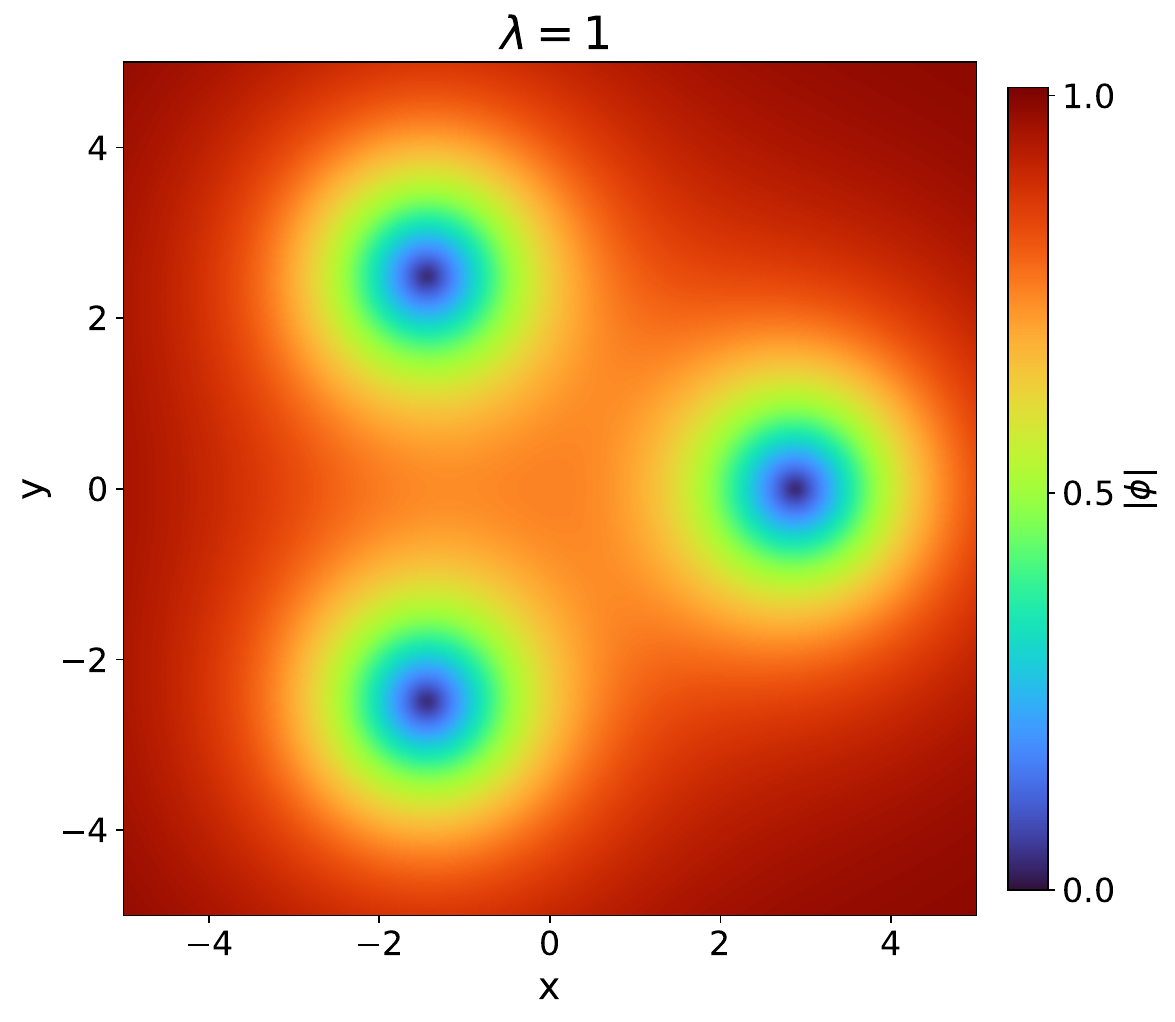}\includegraphics[width=0.376\linewidth]{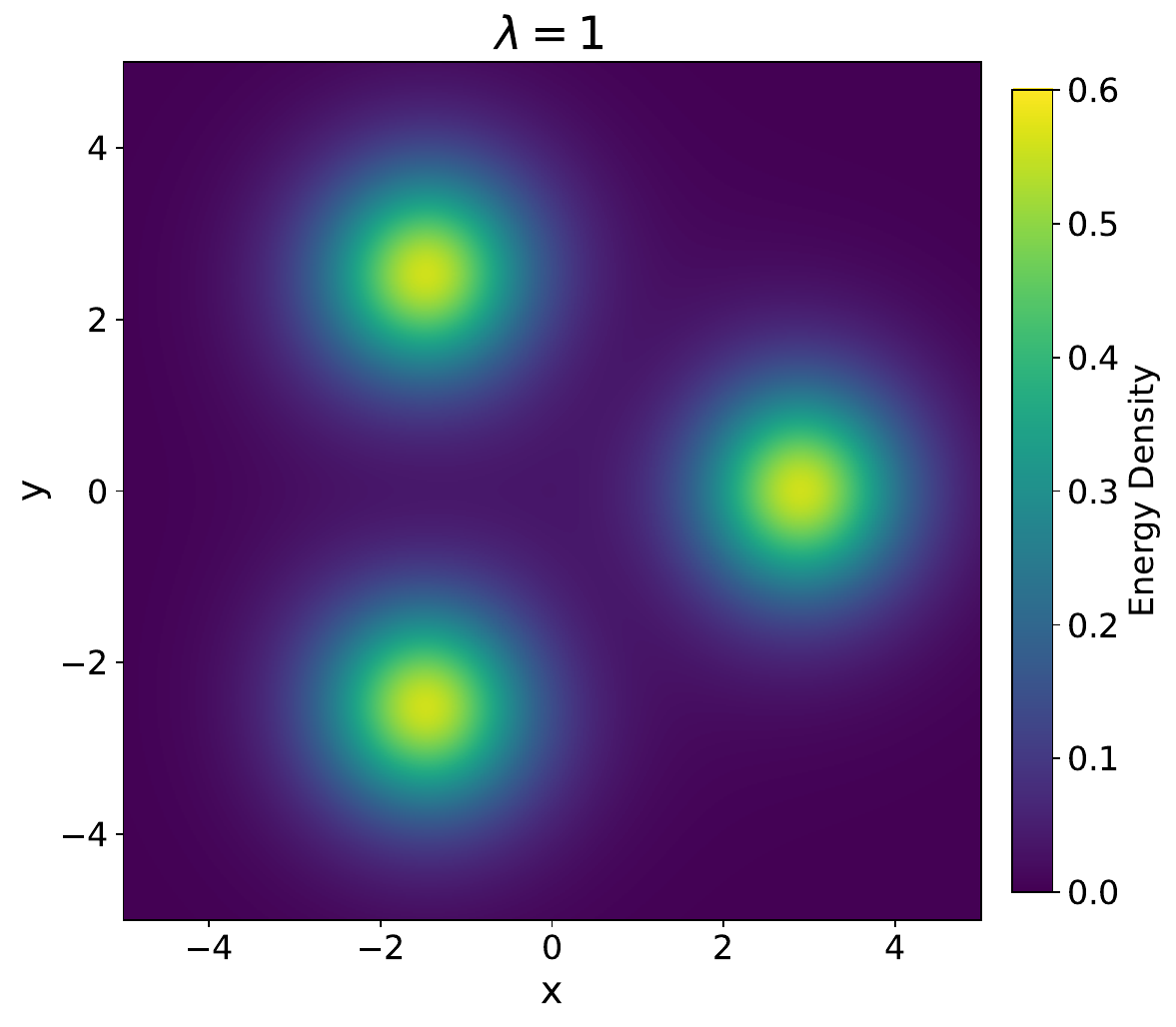}%
        
        \includegraphics[width=0.376\linewidth]{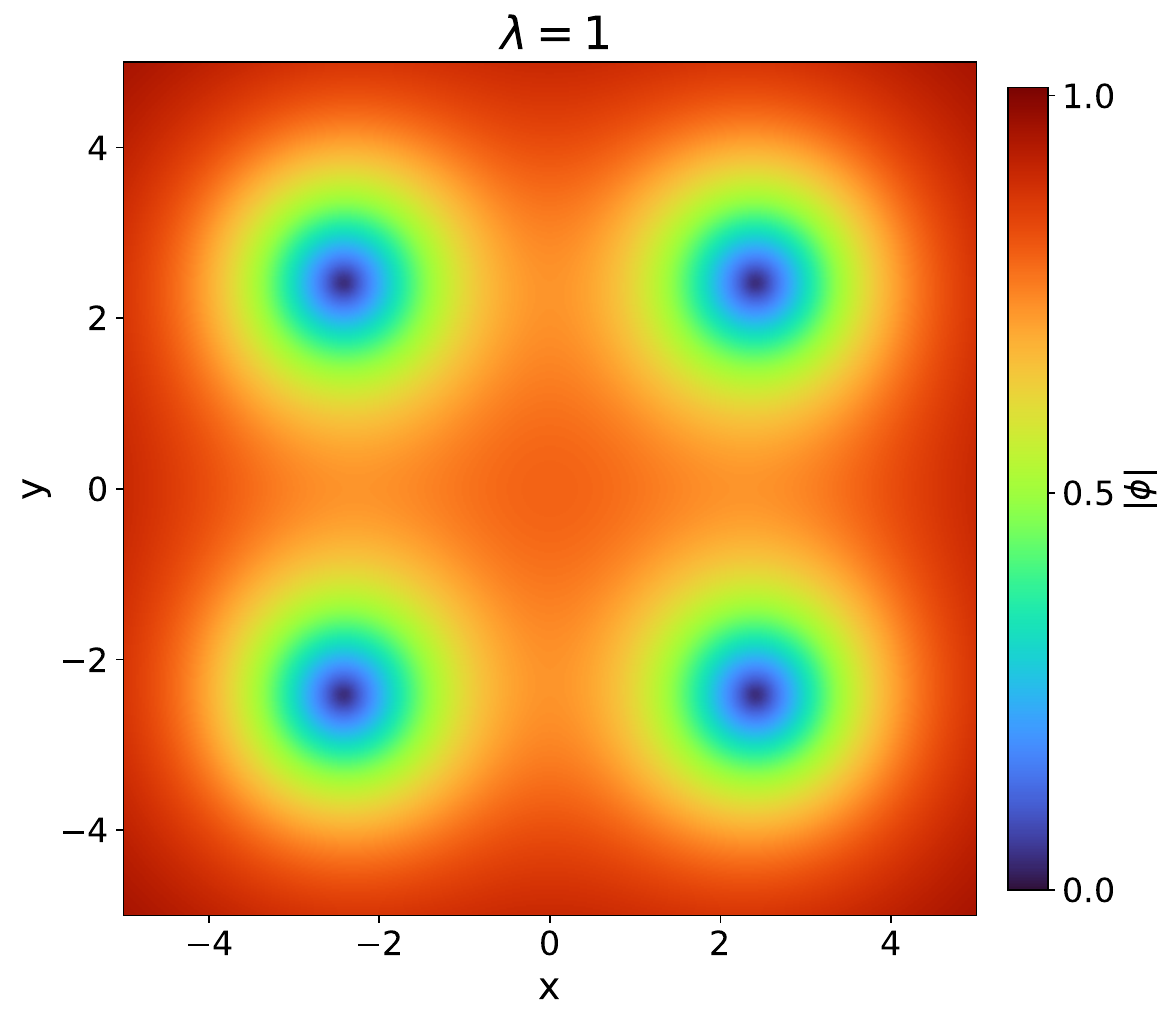}\includegraphics[width=0.376\linewidth]{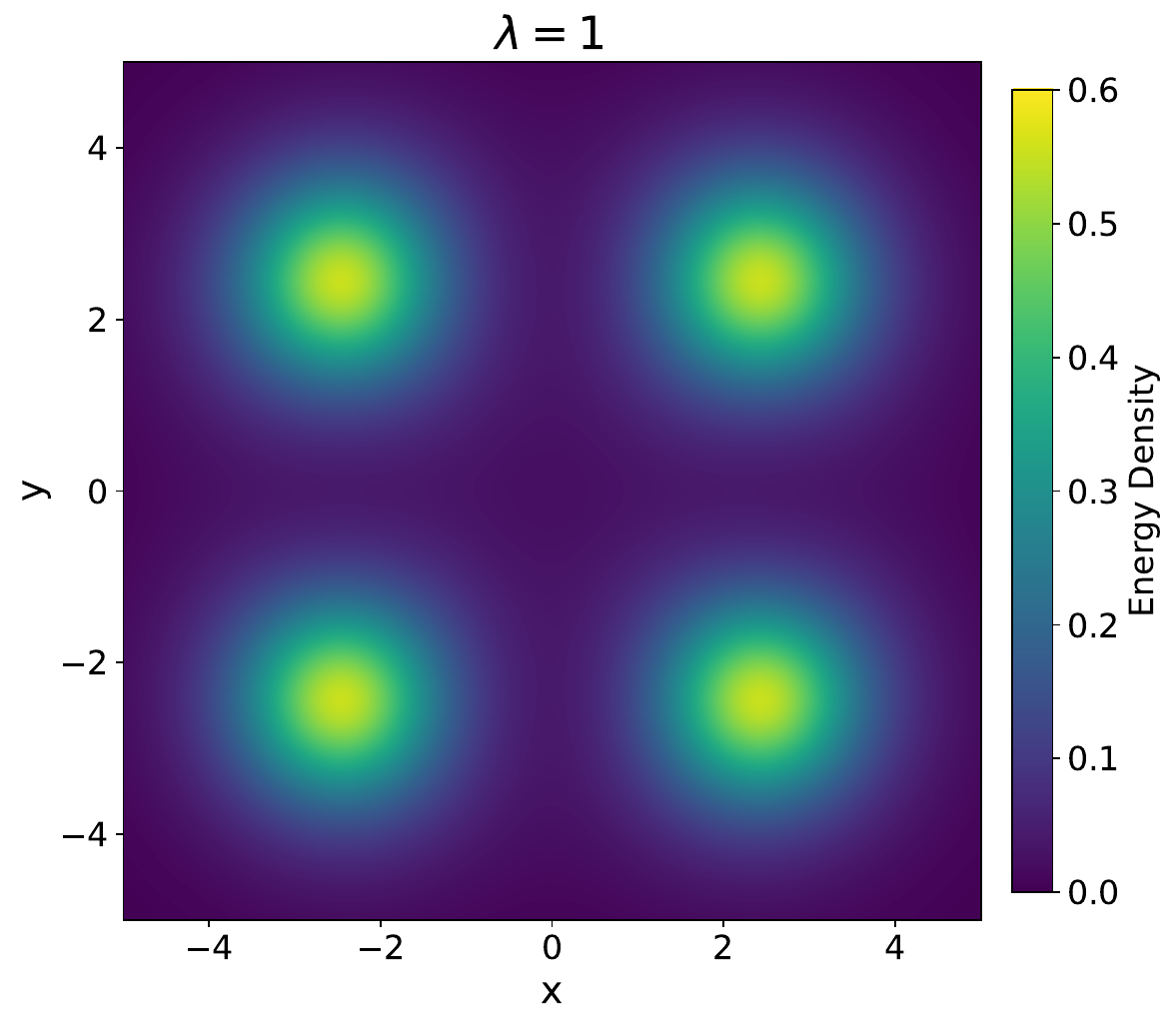}%
    \caption{\small Inner-vortex cluster of winding number $n=3$ (upper panel) and $n = 4$ (lower panel) for $\sigma = 2$ and $\lambda = 1$. We represent the modulus of the scalar field and the energy density.}
    \label{f:Molecule_Inner_Higher_N}
\end{figure}

As expected, in the $n = 2$ case, the transition of the only unstable mode occurs exactly at the first critical value $\lambda_{c1}$ of the self-coupling constant introduced in the previous section. This feature gives us a qualitative idea about when the nonmonotonic interaction appears. In this limit, the mode approaches a zero mode, reflecting the flattening of the interaction energy up to a certain separation. Therefore, vortices can be displaced independently with no change in energy (or negligible) within this range. However, this does not correspond to a genuine BPS limit, because at larger separations a repulsive barrier develops. 

Following this intuition, we have searched for inner-vortex clusters with higher winding numbers. As an illustrative example, we consider for $\sigma=2$ the case $\lambda=1$, for which the system lies sufficiently far from all mode crossings. The modulus of the scalar field and the energy density of the clusters found for winding number $n=3$ and $n = 4$ are shown in Fig. \ref{f:Molecule_Inner_Higher_N}. It is clearly visible that the system adopts a regular polygonal configuration, with the vortices arranged at the vertices of an equilateral triangle for winding number $n=3$ and a square for $n=4$. Notably, the distance between the vortices in the cluster for a given $\lambda$ is not very sensitive to the winding number.  

\begin{figure}[ht!]
    \centering

    \includegraphics[width=0.32\linewidth]{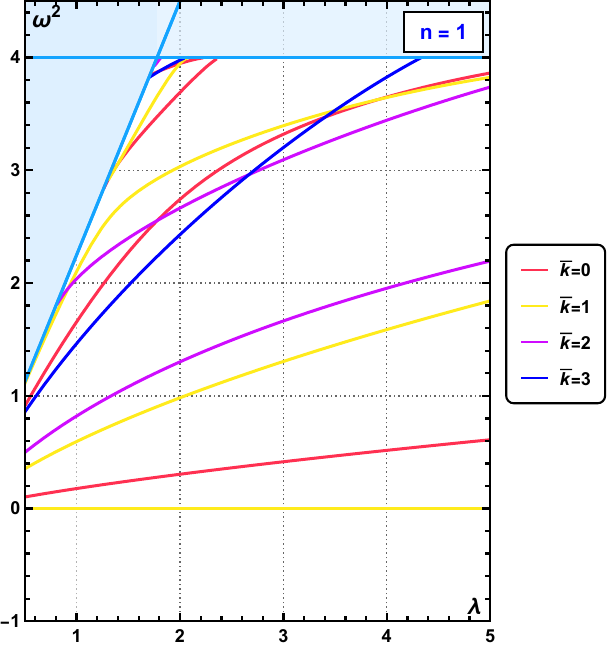}
    \hfill
    \includegraphics[width=0.32\linewidth]{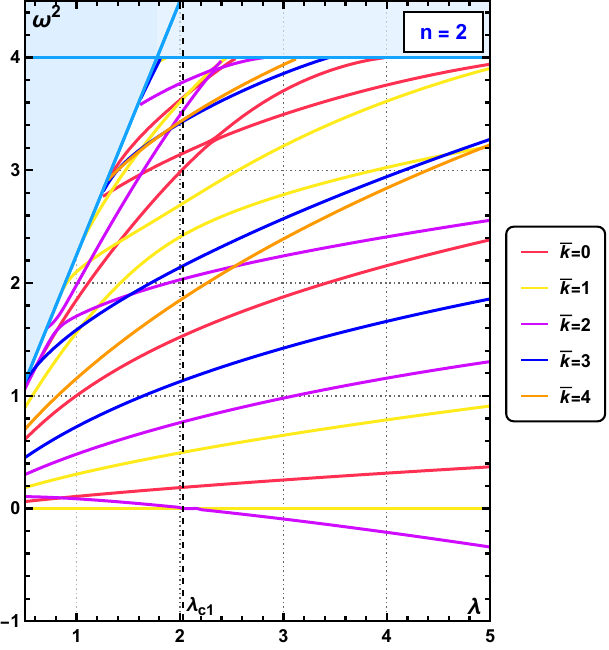}
    \includegraphics[width=0.32\linewidth]{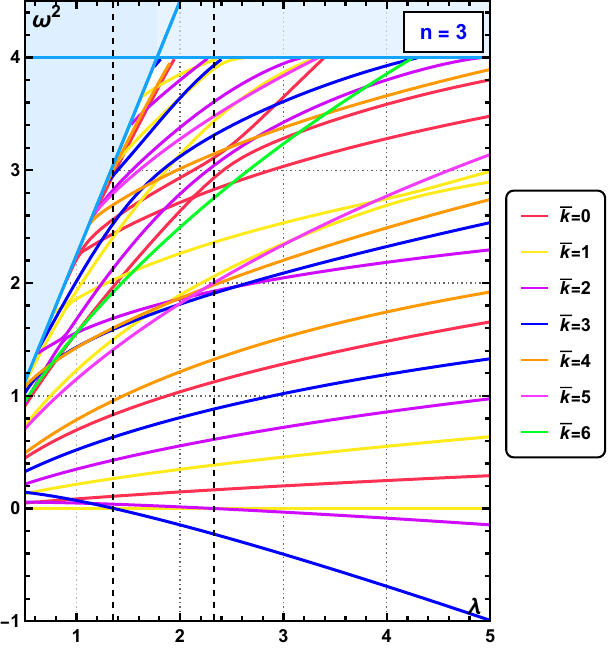}

    \caption{\small Spectral structure for the outer $n$-coincident vortex for $n=1,2$ and $3$. We have assumed $\sigma = 2$ as an illustrative example. The vertical dashed lines indicate the values of $\lambda$ at which multipolar modes becomes unstable.}
    \label{fig:ModesOuter}
\end{figure}

\begin{figure}[ht!]
    \centering
        \centering

        \includegraphics[width=0.4\linewidth]{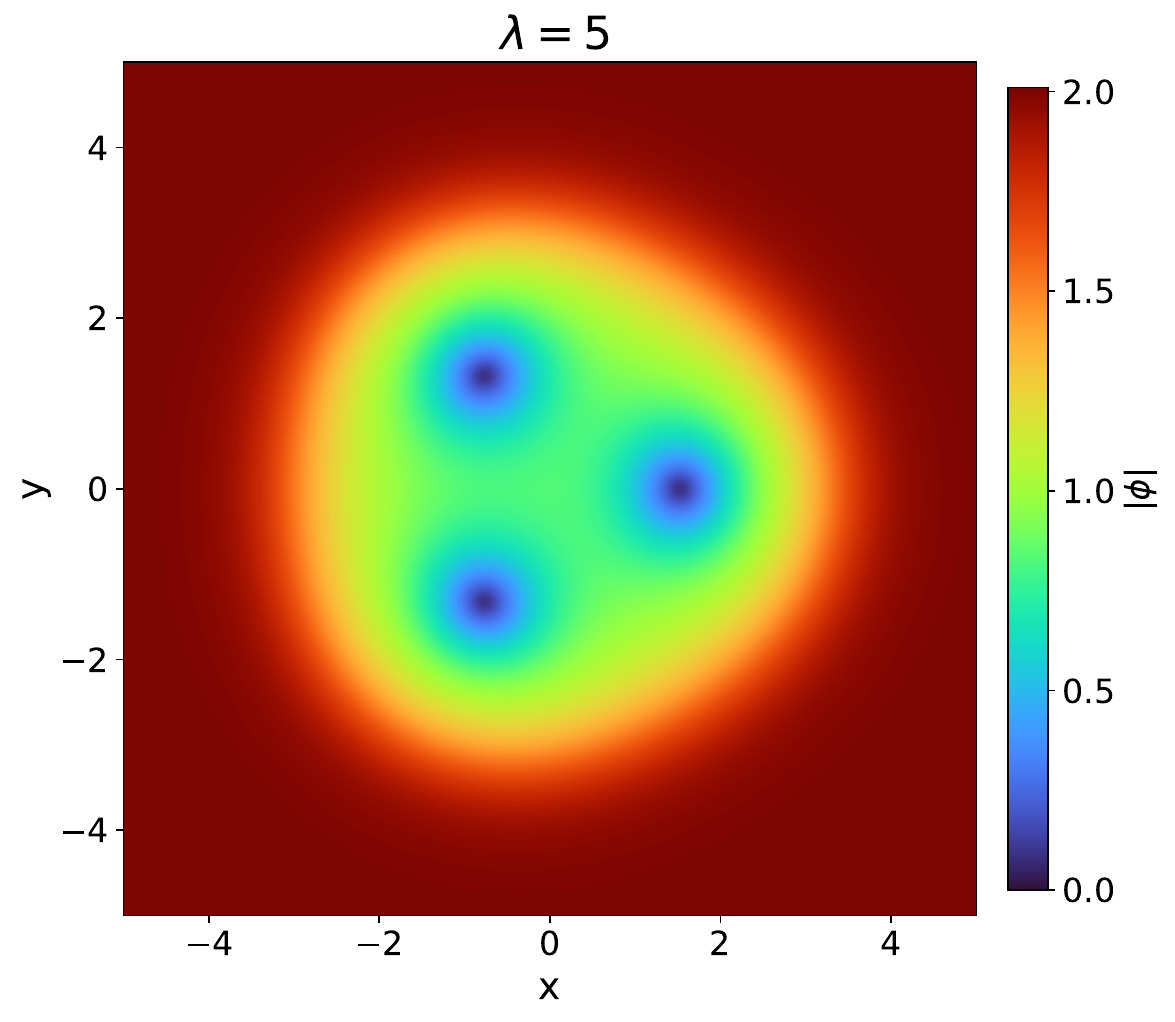}\includegraphics[width=0.4\linewidth]{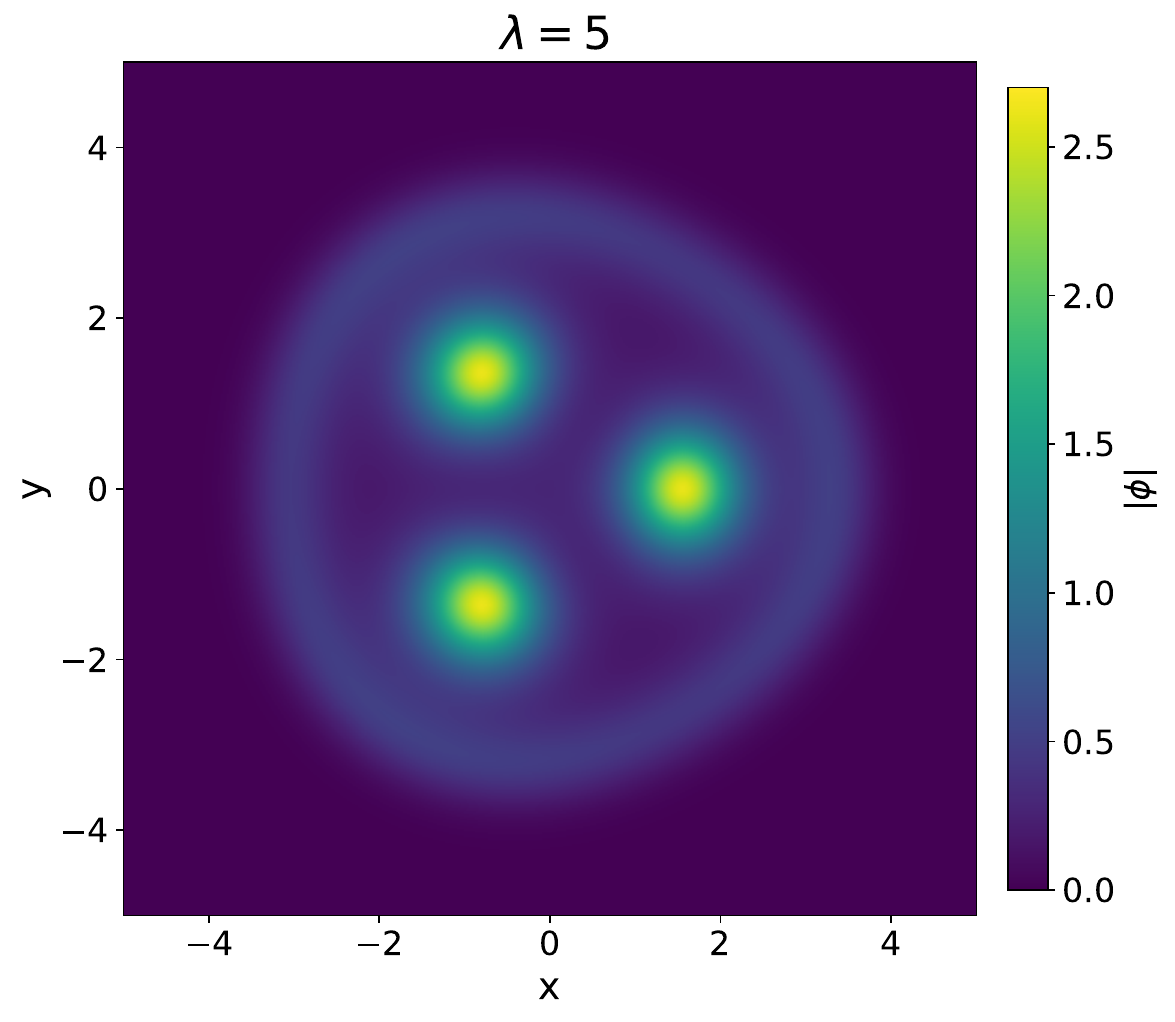}%
 
        \caption{\small Vortex-filled bubble of winding number $n=3$ for $\sigma = 2$ and $\lambda = 5$. We represent the modulus of the scalar field and the energy density.}
    \label{f:Molecule_Outer_Higher_N}
\end{figure}

Regarding the outer-vacuum vortex, its spectral structure is shown in Fig. \ref{fig:ModesOuter} for $\sigma = 2$ and $n = 1,2,3$. Once again, the absence of negative modes confirms the stability of the $1$-vortex. In contrast, the $2$-vortex exhibits a critical value of the self-coupling constant, $\lambda_{c1}$, above which one of the positive modes crosses zero and becomes negative, rendering the vortex unstable. This value coincides with the threshold beyond which we numerically observe the formation of vortex-filled bubbles (see Fig. \ref{f:dVsLambda_Outer}). Using the region where all the unstable modes have crossed the zero frequency axis for $n=3$, we have found a higher winding number bubble. The corresponding configuration is depicted in Fig. \ref{f:Molecule_Outer_Higher_N} for $\sigma = 2$ and $\lambda = 5$.

Remarkably, the outer-vacuum vortex displays a richer internal-mode structure than the inner-vacuum vortex. In particular, outer-vacuum vortices possess multiple Derrick-type modes with $\bar{k}=0$ and a significant number of type-B modes are also present. Furthermore, all these modes cross each other, which suggests that there should appear resonance phenomena among them. This feature is absent for coincident inner-vacuum vortices and for the prototypical axially-symmetric vortices of the standard Abelian–Higgs model. For this reason, the dynamics of outer-vacuum vortices deserves further investigation. However, a detailed analysis lies beyond the scope of the present work and is left for future research.  

\section{Conclusions} \label{s:Conclusions}

In this work, we have analyzed a generalization of the Abelian–Higgs model in which the usual quartic potential is replaced by an eighth-order potential. Specifically, we have considered a model featuring two disconnected continuous vacuum manifolds, corresponding to the circles $|\phi| = 1$ and $|\phi| = \sigma$. Here, $\sigma$ is a parameter that determines the radius of the outer vacuum manifold and encodes the higher-order nonlinearities of the model. This structure gives rise to two distinct species of vortices, as well as domain-wall configurations interpolating between the two vacuum manifolds.

We have determined the profiles of both types of vortices and computed their corresponding energy densities and magnetic fields. A remarkable feature is that outer-vacuum vortices can be interpreted as inner-vacuum vortices surrounded by a domain wall, particularly in the small-$\sigma$ regime. In this regime, the domain-wall tension is low, allowing the wall to spread over a large spatial region. As $\sigma$ increases, the wall tension grows, causing the wall to contract towards the origin and eventually merge with the inner-vacuum vortex. This picture is specially important when the energy interaction between vortices is computed.

Remarkably, our model breaks the conventional type-I and type-II dichotomy of the standard Abelian-Higgs model. In particular, we have found a regime of values of the self-coupling constant in which the interaction energy develops a minimum at a finite separation, suggesting the possibility of stable molecules of vortices in which the vortices remain separated by a characteristic distance. This behavior can be traced back to the higher-order nonlinearities of the potential, which decouple the properties of the vortex cores from those governing the asymptotic tails. This allows the magnetic repulsion to compete with the scalar attraction at intermediate distances. In our case, magnetic interactions dominate at short distances, whereas scalar interactions dominate at large distances.

It is worth emphasizing that the structure of the stable vortex molecules in our model differs significantly between inner- and outer-vacuum vortices. In the case of inner-vacuum vortices, the stable configurations consist of clusters of individual vortices, with the equilibrium separation increasing with the self-coupling constant $\lambda$. By contrast, stable molecules of outer-vacuum vortices take the form of vortex-filled bubbles, in which the vortex cores, corresponding to inner-vacuum vortices, are surrounded by a domain wall. Moreover, the separation between the vortex cores within the bubble approaches a constant value in the large-$\lambda$ limit. It would be interesting to investigate the modifications of these results when one of the vacua is lifted. 

Another striking difference concerning the outer-vacuum vortices is their phase transitions when they interact with other outer-vacuum vortices or with a domain wall. This phase transition is identified by discontinuous jumps observed in its interaction energy with these objects. This behavior can be attributed to the role of the domain wall constituting the outer-vacuum vortex. At large separations, outer-vacuum vortices interact only weakly with other outer-vacuum vortices or domain walls, owing to the exponential decay of their asymptotic tails. However, as the separation decreases, an outer-vacuum vortex can lose its identity as an individual soliton, and the interaction energy becomes dominated by the interaction between the associated wall and the other solitonic object. This change in the underlying configuration gives rise to a discontinuity in the interaction energy. By contrast, the interaction energy of inner-vacuum vortices varies smoothly in all these settings. A possible extension of the present work is to determine whether the phase transitions observed in the interactions between outer-vacuum vortices and between an outer-vacuum vortex and a domain wall are first-order transitions. To this end, one could investigate whether multiple branches of solutions coexist.

Finally, we have analyzed the linear spectrum of perturbations around coincident $n$-vortices, identifying their unstable modes and positive-frequency bound modes. Unlike in the standard Abelian–Higgs model, we find that Type-A multipolar modes of both inner- and outer-vacuum vortices become unstable at different values of $\lambda$, rather than becoming degenerate at $\omega^2=0$ at a common value. In particular, for the coincident $n=2$ vortex, the zero crossing of the only unstable mode identifies the lower threshold where the interaction energy exhibits a minimum. Another distinctive feature of our model is that inner-vacuum vortices do not support Type-B modes. More interestingly, outer-vacuum vortices support a large number of Derrick-type and Type-B modes that cross each other. This degeneration can lead to resonances that drastically change the dynamics. It would therefore be interesting to investigate the dynamics of the different vortex species and their possible scattering outcomes. This analysis is left for future work. 

\begin{figure}[ht!]
    \centering
        \centering
        \hspace{0.3cm}
        \includegraphics[width=0.379\linewidth]{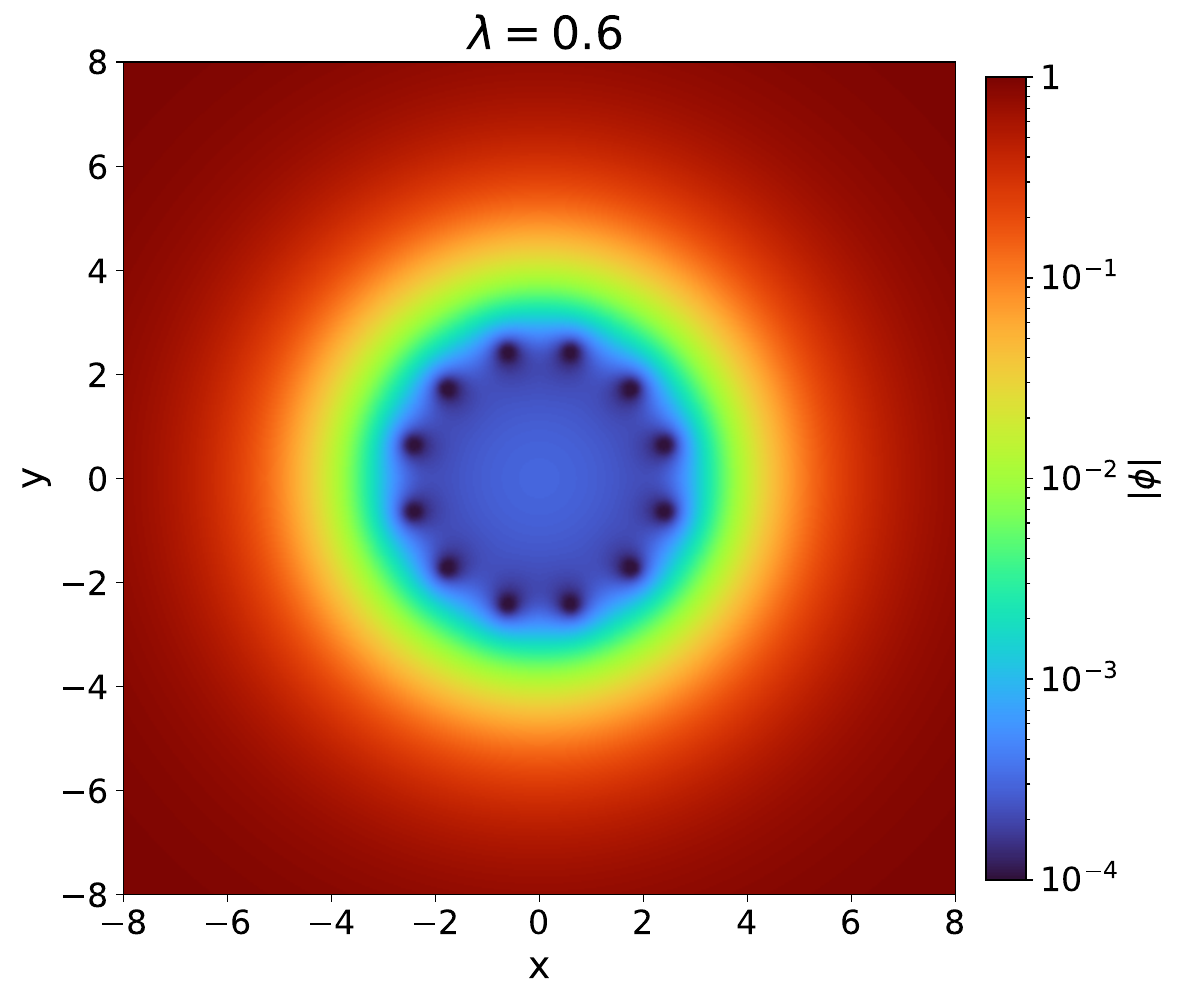}        \includegraphics[width=0.374\linewidth]{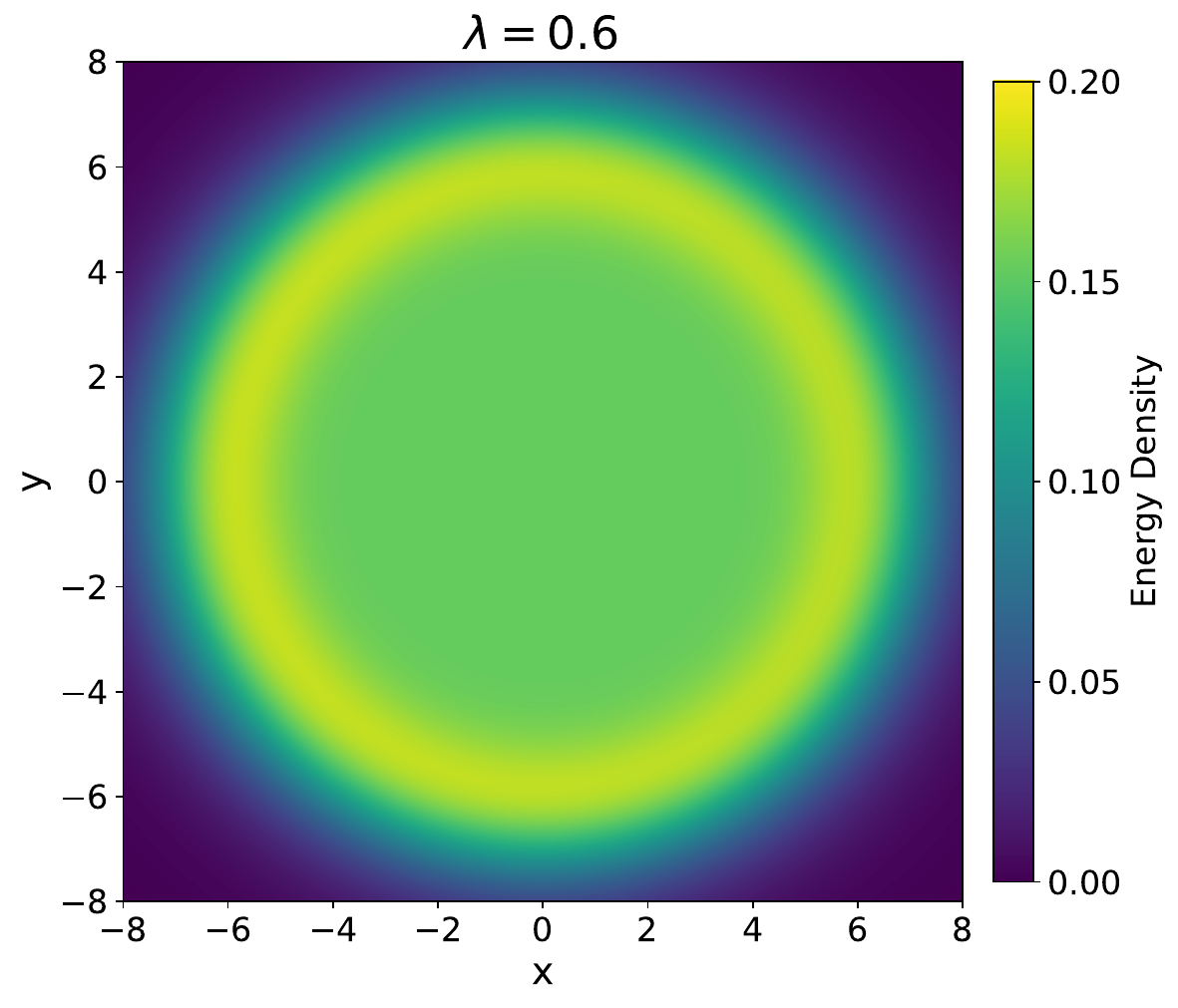}
        
        \includegraphics[width=0.39\linewidth]{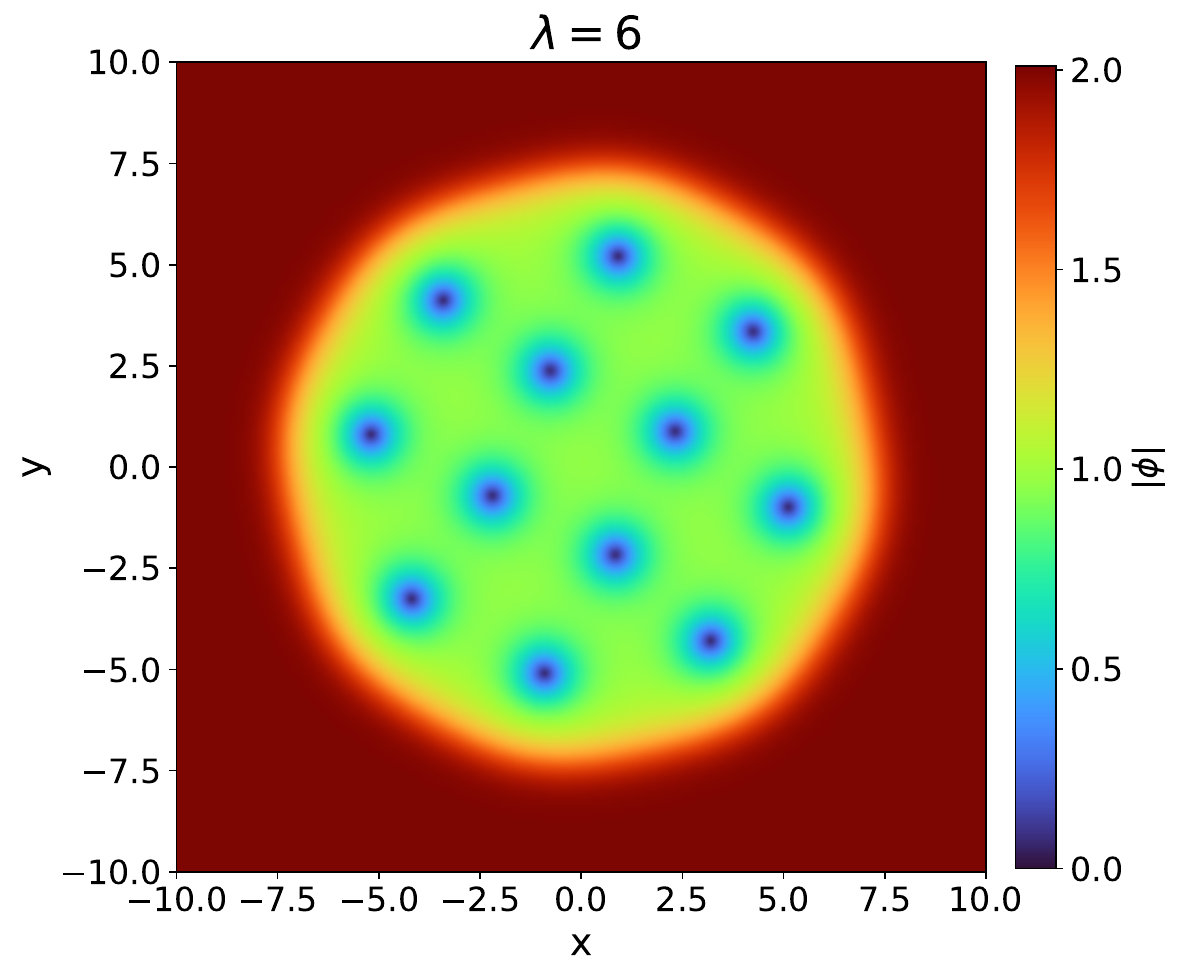}\includegraphics[width=0.39\linewidth]{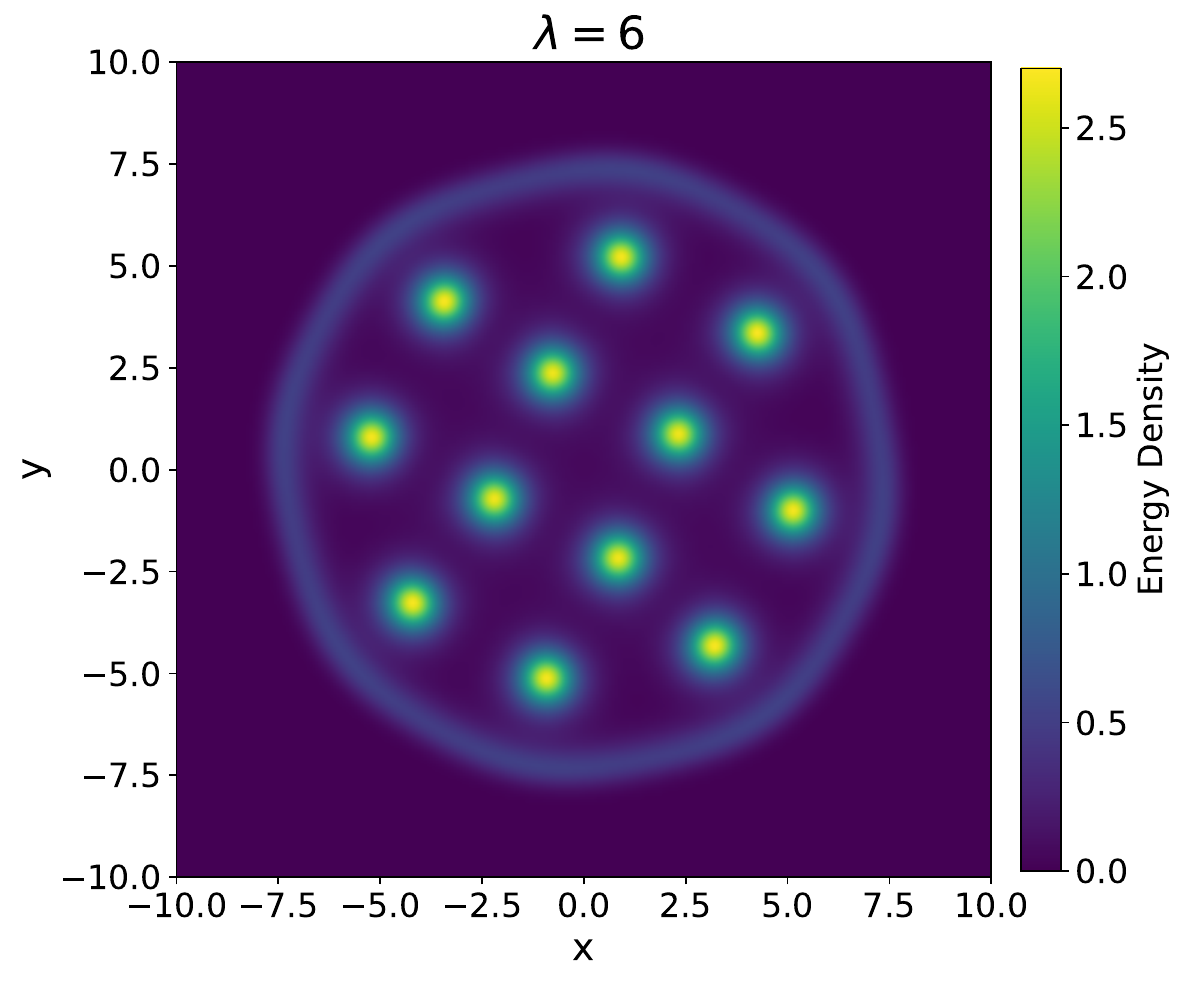}%
 
        \caption{\small Modulus of the scalar field and energy density of the inner-vortex cluster (upper panel) and the vortex-filled bubble (lower panel) of winding number $n = 12$ for $\sigma = 2.0$.}
    \label{f:Molecule_High_N}
\end{figure}

The high-winding-number regime also deserves special attention. As shown in Fig. \ref{f:Molecule_High_N}, we can find stable vortex molecules even for arbitrary high vorticities. The outer-vacuum configuration still takes the form of a vortex-filled bubble, in which individual vortices are enclosed by a domain wall. By contrast, inner-vacuum vortices arrange themselves into regular polygons. Nevertheless, we expect that the geometries of the stable configurations are sensitive to the value of $\lambda$ considered, due to the increasing number of unstable modes and the interactions among multiple vortices.

Although we have investigated the Abelian-Higgs model with a modified potential, it is possible that similar results can be obtained in more general theories. For example, further possible extensions of the present work include
supersymmetric theories \cite{DiVecchia:1977nxl,Witten:1978mh,Edelstein:1993bb,Tong:2005un,Eto:2006pg,Shifman:2007ce,Shifman:2009zz,Tong:2008qd}, 
non-Abelian vortices 
in the non-Abelian Higgs model 
\cite{Hanany:2003hp,Auzzi:2003fs,Eto:2004rz,Eto:2005yh,Eto:2006cx,Eto:2006uw,Eto:2006db,Tong:2005un,Eto:2006pg,Shifman:2007ce,Shifman:2009zz,Tong:2008qd} or 
semi-local vortices 
with  additional Higgs fields 
\cite{Vachaspati:1991dz,Achucarro:1999it,Shifman:2006kd,Eto:2007yv}.  

\section*{Acknowledgments}

D.M.C. and S.N.O. have been supported in part by the grant PID2023-148409NB-I00 MTM. D.M.C. and S.N.O. also acknowledge financial support from the European Social Fund, the Operational Programme of the Junta de Castilla y Le\'on, and the Regional Ministry of Education. Financial support from the Department of Education of the Junta de Castilla y León and FEDER Funds is also gratefully acknowledged through the grant CLU-2023-1-05. M.N. acknowledges the program ``Sustainability with Knotted Chiral Meta Matter (WPI-SKCM$^2$)'' at Hiroshima University and JSPS KAKENHI [Grants No. JP22H01221, JP23K22492].

\section*{Conflict of Interest}
The authors declare that they have no conflict of interest.
 
\section*{Data availibility}
Data sharing is not applicable to this article as no datasets were generated or analysed during
the current study.

\printbibliography[heading=bibintoc, title={References}]
\end{document}